\documentclass[12pt]{article}
\usepackage{amssymb,epsf,color}
\def\lsim{\mathrel{\rlap {\raise.5ex\hbox{$ < $}}
{\lower.5ex\hbox{$\sim$}}}}
\def\gsim{\mathrel{\rlap {\raise.5ex\hbox{$ > $}}
{\lower.5ex\hbox{$\sim$}}}} 
\def\sqr#1#2{{\vcenter{\vbox{\hrule height.#2pt

        \hbox{\vrule width.#2pt height#1pt \kern#1pt

           \vrule width.#2pt}

        \hrule height.#2pt}}}}

\def\lsim{{\displaystyle
{{\raise-8pt\hbox{$ <$}}
\atop{\raise5pt\hbox{$\sim$}}}}}
\def\gsim{{\displaystyle
{{\raise-8pt\hbox{$ >$}}
\atop{\raise5pt\hbox{$\sim$}}}}}
\def\slsim{{\displaystyle
{{\raise-8pt\hbox{$\scriptstyle <$}}
\atop{\raise5pt\hbox{$\scriptstyle \sim$}}}}}
\def\sgsim{{\displaystyle
{{\raise-8pt\hbox{$\scriptstyle  >$}}

\atop{\raise5pt\hbox{$\scriptstyle \sim$}}}}}

\newskip\humongous \humongous=0pt plus 1000pt minus 1000pt

\newcommand{\sumpf}[0]{\sum_{(H^{\rm f},G^{\rm f})}\! \! \! \!
{\raise
4pt
\hbox{$'$}}\,}

\newcommand{\sump}[0]{\sum_{(H,G)}\! \! {\raise 4pt \hbox{$'$}}\,}

\def\bs{\begin{subequations}}
\def\es{\end{subequations}}

\catcode`\@=11
\newcount\hour
\newcount\minute
\newtoks\amorpm

\hour=\time\divide\hour by60
\minute=\time{\multiply\hour by60 \global\advance\minute by-\hour}
\edef\standardtime{{\ifnum\hour<12 \global\amorpm={am}%
        \else\global\amorpm={pm}\advance\hour by-12 \fi

        \ifnum\hour=0 \hour=12 \fi
        \number\hour:\ifnum\minute<10 0\fi\number\minute\the\amorpm}}
\edef\militarytime{\number\hour:\ifnum\minute<10 0\fi\number\minute}
\def\draftlabel#1{{\@bsphack\if@filesw {\let\thepage\relax
   \xdef\@gtempa{\write\@auxout{\string
      \newlabel{#1}{{\@currentlabel}{\thepage}}}}}\@gtempa
   \if@nobreak \ifvmode\nobreak\fi\fi\fi\@esphack}
        \gdef\@eqnlabel{#1}}
\def\@eqnlabel{}
\def\@vacuum{}
\def\draftmarginnote#1{\marginpar{\raggedright\scriptsize\tt#1}}
\def\draft{\oddsidemargin -.2truein
        \def\@oddfoot{\sl preliminary draft \hfil
        \rm\thepage\hfil\sl\today\quad\militarytime}
        \let\@evenfoot\@oddfoot \overfullrule 3pt
        \let\label=\draftlabel
        \let\marginnote=\draftmarginnote
   \def\@eqnnum{(\theequation)\rlap{\kern\marginparsep\tt\@eqnlabel}%
\global\let\@eqnlabel\@vacuum}  }
\def\subequations{\refstepcounter{equation}%
  \edef\@savedequation{\the\c@equation}%
  \@stequation=\expandafter{\theequation}
  \edef\@savedtheequation{\the\@stequation}
  \edef\oldtheequation{\theequation}%
  \setcounter{equation}{0}%
  \def\theequation{\oldtheequation\alph{equation}}}

\def\endsubequations{\setcounter{equation}{\@savedequation}%
  \@stequation=\expandafter{\@savedtheequation}%
  \edef\theequation{\the\@stequation}\global\@ignoretrue
  \vspace*{-12pt} \\}

\def\bs{\begin{subequations}}
\def\es{\end{subequations}}
\relax
\def\thefootnote{\fnsymbol{footnote}}
\def\be{\begin{equation}}
\def\ee{\end{equation}}
\def\ba{\begin{eqnarray}}
\def\ea{\end{eqnarray}}

\def\ee{\end{equation}}
\def\bea{\begin{eqnarray}}
\def\eea{\end{eqnarray}}
\def\nn{\nonumber}

\newcommand{\uarrw}[0]{\mathrel{
{\raise.5ex\vbox{\hrule width 1cm}\hskip-6pt\rightarrow}}}
\def\thebibliography#1{%
\vskip 0.5cm \centerline{\bf References}
\list{%
[\arabic{enumi}]}{\settowidth\labelwidth{[#1]}
\leftmargin\labelwidth
\advance\leftmargin\labelsep
\usecounter{enumi}}
\def\newblock{\hskip .11em plus .33em minus .07em}
\sloppy\clubpenalty4000\widowpenalty4000
\sfcode`\.=1000\relax}

\renewcommand{\theequation}{\arabic{section}.\arabic{equation}}

\renewcommand{\section}{\setcounter{equation}{0}\@startsection%
{section}{1}{0mm}{-\baselineskip}{0.5\baselineskip}%
{\normalfont\normalsize\bfseries}}

\renewcommand{\subsection}{\@startsection%
{subsection}{2}{0mm}{-\baselineskip}{0.5\baselineskip}%
{\normalfont\normalsize\slshape}}

\renewcommand{\subsubsection}{\@startsection%
{subsubsection}{2}{0mm}{-\baselineskip}{0.5\baselineskip}%
{\normalfont\normalsize\slshape}}

\begin{document}

\renewcommand{\theequation}{\arabic{section}.\arabic{equation}}
\begin{titlepage}
\begin{flushright}
\end{flushright}
\begin{centering}
\vspace{1.0in}
\boldmath

{ \large \bf A physical universe from the universe of codes} 

\unboldmath
\vspace{2cm}

{\bf Andrea Gregori}$^{\dagger}$ \\
\medskip
\vspace{3cm}
{\bf Abstract} \\
\end{centering} 
\vspace{.2in}
We investigate the most general phase space of configurations, consisting 
of the collection of all possible ways of assigning elementary attributes, 
``energies'', to elementary positions, ``cells''. 
We discuss how this space defines a ``universe'' with a structure that 
can be approximately described by a quantum-relativistic physical scenario
in three space dimensions.
In particular, we discuss how the Heisenberg's 
Uncertainty and the bound on the speed of light arise, and 
what kind of mechanics rules on this space.

\vspace{8cm}

\hrule width 6.7cm
\noindent
$^{\dagger}$e-mail: agregori@libero.it

\end{titlepage}
\newpage
\setcounter{footnote}{0}
\renewcommand{\thefootnote}{\arabic{footnote}}

\vspace{1.5cm} 

\tableofcontents

\vspace{1.5cm}

\section{Introduction}
\label{intro}

The search for a unified description of 
quantum mechanics and general relativity, within a theory that should
possibly describe also the evolution of the universe, is one of the long 
standing and debated open problems of modern theoretical physics. 
The hope is that, once such a theory has been found, it will
open us a new perspective from which to approach,
if not really answer, 
the fundamental question behind all that, that is ``why the universe is what
it is''. On the other hand, it is not automatic that, once such a unified
theory has been found, it gives us also more insight on the reasons
why the theory is what it is, namely, \emph{why} it has to be precisely 
that one, and why no other choice could work. But perhaps it is precisely
going first through this question that it is possible to make progress in
trying to solve the starting problem, namely the one of unifying quantum 
mechanics and relativity. Indeed, after all we don't know why do we need
quantum mechanics, and relativity, or, equivalently, why the speed
of light is a universal constant, or why there is the Heisenberg Uncertainty.
We simply know that, in a certain regime, Quantum Mechanics and Relativity
work well in describing physical phenomena.

In this work, we approach the problem from a different perspective.
We do not assume quantum mechanics, nor relativity.
The question we start with can be formulated as follows: is it possible
that the physical world, as we see it, doesn't proceed from a
``selection'' principle, whatever
this can be, but it is just the collection of all the
possible ``configurations'', intended in the most
general meaning? May the history of the Universe be viewed somehow as a
path through these configurations, and what we call time ordering an
ordering through the inclusion of sets, so that the universe at a
certain time is characterized by its containing as subsets all previous
configurations, whereas configurations which are not 
contained belong to the future of the Universe? What is the
meaning of ``configuration'', and
how are then characterized configurations, in order to say which one is
contained and which not? How do they contribute to build up what we observe?

Let us consider the most general 
possible phase space of ``spaces of codes of information''.
By this we mean products of spaces carrying strings of information
of the type ``1'' or ``0'' (we will comment at the end of the paper about the
generality of the choice of working with binary codes). 
If we interpret 
these as occupation numbers for
cells that may bear or not a unit of energy, we can view the set of these
codes as the set of assignments of a map $\Psi$ from a space of unit
energy cells to a discrete target vector space, that can be of any 
dimensionality. 
If we appropriately introduce units of length and energy, we may ask
what is the geometry of any of these spaces. 
Once provided with this interpretation, it is clear that the problem of
classifying all possible information codes can be viewed as a classification
of the possible geometries of space, of any possible dimension.
If we consider the set of all
these spaces, i.e. the set of all maps, $\left\{ \Psi \right\}$,  
that we call the phase space of all maps, we may also ask whether 
some geometries occur more or less often in this phase space. 
In particular, we may ask this question about $\left\{ \Psi (E) \right\}$,
the set of all maps which assign a finite amount of energy
units, $N \equiv E$. The frequency by which these spaces occur
depends on the combinatorics of the energy assignments 
\footnote{In order to unambiguously define these frequencies, 
it is necessary to make a ``regularization''
of the phase space by imposing to work at finite volume. This 
condition can then be relaxed once a regularization-independent
prescription for the computation of observables is introduced.}. 
Indeed, it turns out that not only 
there are configurations which occur more often than other ones, but
that there are no two configurations with the same weight.
If we call $\left\{ \Psi (E) \right\}$ the ``universe'' at ``energy'' $E$,
we can see that we can assign a time ordering in a natural way, because
$\left\{ \Psi (E^{\prime}) \right\}$ ``contains''
$\left\{ \Psi (E) \right\}$ if
$E^{\prime } > E$, in the sense that 
$\forall \Psi \in \left\{ \Psi (E) \right\}$ $\exists \Psi^{\prime}
\in \left\{ \Psi (E^{\prime}) \right\}$ such that
$\Psi \subsetneqq \Psi^{\prime}$. $E$ plays therefore the role of 
a time parameter, that we can call the age of the universe, ${\cal T}$. 
Our fundamental assumption is that, at any time $E$, 
there is no ``selected'' geometry of the universe: the universe as it
appears is given by the superposition of all possible geometries.
Namely, we assume that
the partition function of the universe, i.e. the function through which
all observables are computed, is given by:
\be
{\cal Z} (E) \, = \, \sum_{\Psi(E)} {\rm e}^{S(\Psi(E))} \, ,
\label{zsum1}
\ee 
where $S(\Psi)$ is the entropy of the configuration $\Psi$ in the phase space
$\left\{  \Psi \right\}$, related to the weight of occupation in the
phase space $W(\Psi)$ in the usual way: $S = \log W$. Rather evidently,
the sum is dominated by the configurations of
highest entropy.
The most recurrent geometries 
of this universe turn out to be those corresponding to three dimensions.
Not only, but the very dominant configuration is the one that,
in the continuum limit, corresponds
to a three-sphere of radius $R$ proportional to $E$. 
That is, a black hole-like 
universe in which the energy density 
is $\sim 1 / E^2 \propto 1 / R^2$~~\footnote{The radius of the black hole
is the radius of the three-ball enclosed by the horizon surface. 
The radius of the three sphere does not coincide with the radius of the ball; 
they are anyway proportional to each other. How,
and in which sense, a sphere can be said to have, like a ball, 
a boundary, which works as horizon, is a rather non-trivial 
fact related to the very special topology of this space,
discussed in detail in Ref.~\cite{assiom-2011}.}.
In this scenario there is basically no free
parameter except for the only running quantity, the age of the
universe, in terms of which everything is computed.
Out of the dominant configuration, a three-sphere,
the contribution given by the other configurations to (\ref{zsum1})
is responsible for the introduction of ``inhomogeneities'' in the universe. 
These are what gives rise to a varied spectrum of energy clusters, 
that we interpret as matter and fields evolving and interacting
during a time evolution set by the $E$--time-ordering.

The most striking feature is that all these configurations 
summed up contribute
for a correction to the total energy of the universe of the 
order of $\Delta E \sim 1 / {\cal T}$. 
This is rather reminiscent of the inequality at the base of the
Heisenberg Uncertainty Principle on which quantum mechanics is based on:
${\cal T}$, the age/radius up to the horizon of observation,
can also be written as $\Delta t$, the interval of time during which
the universe of radius $E$ has been produced. That means, the universe is
mostly a classical space, plus a
``smearing'' that quantitatively
corresponds to the Heisenberg uncertainty, $\Delta E \sim 1 / \Delta t$.
This argument can be refined and applied to any observable one may
define: all what we observe is given by a superposition of
configurations and whatever value of observable quantity we can
measure is smeared around, is given with a certain fuzziness, which
corresponds to the Heisenberg's inequality.
Indeed, a more detailed inspection
of the geometries that arise in this scenario, the way ``energy clusters''
arise, their possible interpretation in terms of matter, particles etc.
allows to conclude that \ref{zsum1} formally implies a quantum
scenario, in which the Heisenberg Uncertainty
receives a \emph{new interpretation}.
The Heisenberg uncertainty relation arises here as a
way of accounting not
simply for our ignorance about the observables, but for the 
ill-definedness of these quantities in themselves: all the observables
that we may refer to a three-dimensional world, together with
the three-dimensional space itself, exist only
as ``large scale'' effects. Beyond a certain degree of accuracy
they can neither be measured nor be defined. The space itself, with
a well defined dimension and geometry, 
cannot be defined beyond a certain degree of accuracy either. 
This is due to the fact that
the universe is not just given by one configuration, the dominant one,
but by the superposition of all possible configurations, an infinite
number, among which many (an infinite number too) don't even correspond to a
three dimensional geometry.

It is possible to show that the speed of expansion of the 
geometry of the dominant configuration of the universe,
i.e. the speed of expansion of the radius of the three-dimensional black hole, 
that by convention and choice of
units we can call ``$c$'', is also the
maximal speed of propagation of \underline{coherent}, i.e.
non-dispersive, information. This can be shown
to correspond to the bound 
of the speed of light (see Ref.~\cite{assiom-2011}).
Here it is
essential that we are talking of \emph{coherent information}, as tachyonic
configurations also exist and contribute to~\ref{zsum1}: their contribution
is collected under the Heisenberg uncertainty. 
One may also show that
the geometry of geodesics in this space corresponds to
the one generated by the energy distribution. All this means that this
framework ``embeds'' in itself special and general relativity.

The dynamics implied by (\ref{zsum1}) is neither deterministic in the
ordinary sense of causal evolution, nor
probabilistic. At any age the universe is the superposition
of all possible configurations, weighted by their ``combinatorial''
entropy in the phase space. According to our definition of time and time
ordering,
at any time the actual superposition of configurations does not depend on
the superposition at a previous time, because the actual and the previous one
trivially are 
the superposition of all the possible configurations at their time. 
Nevertheless,
on the large scale the flow of mean values through the time
can be approximated by a smooth evolution that we can, up to a certain extent,
parametrize through evolution equations. As it is not possible
to exactly perform the sum of infinite terms of \ref{zsum1},
and it does not even make sense, because an infinite number of less
entropic configurations don't even correspond to a description of
the world in terms of three dimensions,
it turns out to be convenient to
accept for practical purposes
a certain amount of unpredictability, introduce probability
amplitudes and work in terms of the rules of quantum mechanics. These
appear as precisely tuned to embed the uncertainty that
we formally identified with the Heisenberg Uncertainty
into a viable framework, which allows some
control of the unknown, by endowing the uncertainty with a probabilistic
interpretation.
Within this theoretical framework, we can therefore give an
argument for the \underline{necessity} of a quantum description of
the world: quantization appears to be a useful
way of parametrizing the fact of being the observed reality a
superposition of an infinite number of configurations.
Once endowed with this interpretation,
this scenario provides us with a theoretical framework that
unifies quantum mechanics and relativity in a description that,
basically, is neither of them: in this perspective, they turn out to be only
approximations, valid in a certain limit, of a more comprehensive
formulation.

As discussed in \cite{assiom-2011}, the ``spectrum'' of the theory, namely
the microscopical content of particles and their interactions,
can be investigated via string theory tools. In this framework,
String Theory arises as a consistent quantum theory
of gravity and interacting fields and particles, 
which constitutes a useful mapping of the
combinatorial problem of ``distribution of energy
along a target space'' into a continuum space. Once so interpreted, 
it is no more a ``free'' theory. Like the physics
implied by \ref{zsum1}, it is on the contrary highly predictive. 
Within this framework it is
even possible to see its uniqueness \cite{assiom-2011}.  
For a detailed analysis of the spectrum of the theory implied
by \ref{zsum1}, and the phenomenological implications,
we refer the reader to \cite{npstrings-2011}, \cite{spc-gregori}
\cite{blackholes-g}, and~\cite{paleo}. In particular, Refs.~\cite{spc-gregori}
and~\cite{paleo} show how this theoretical framework, being on its ground a 
new approach to quantum mechanics and phenomenology, does not simply provide us
with possible answers to problems which are traditionally referred to 
quantum gravity and string theory, like the computation of spectrum, masses
and couplings of the elementary particles, but opens new perspectives about 
problems apparently
pertaining to other domains of physics, such
as (high temperature) superconductivity and evolutionary biology.

\section{The set-up}
\label{setup}

\subsection{Distributing degrees of freedom}
\label{ddf}

Consider a generic ``multi-dimensional'' space, consisting of 
$M_1^{p_1} \times \ldots \times M_i^{p_i} \ldots \times M_n^{p_n}$ 
``elementary cells''. Since an elementary, ``unit'' 
cell is basically a-dimensional, it makes sense to measure the volume
of this $p$-dimensional space, $p = \sum^n_i p_i$, in terms of unit cells:
$V = M_1^{p_i} \times \ldots \times M_n^{p_n}$. 
Although with the same volume, from the point of 
view of the combinatorics of cells and attributes this space is deeply
different from a one-dimensional space with $V$ cells. 
However, independently on the dimensionality, 
to such a space we can in any case assign, in the sense of ``distribute'',
$N$ ``elementary'' attributes, $N \leq V$. Indeed, in order to preserve the
basic interpretation of the ``$N$'' coordinate as ``attributes'' and
the ``$M$'' degrees of freedom as ``space'' coordinates, to which attributes
are assigned, it is necessary that $N \leq M_n$, $\forall \, n$ 
\footnote{In the case 
$N > M_n$ for some $n$, we must interchange the interpretation of 
the $N$ as attributes and instead consider them as a space coordinate, whereas
it is $M_n$ that are going to be seen as a coordinate of attributes.}.   
What are these attributes? Cells, simply cells:
our space is simply a mathematical
structure of cells, and cells that we attribute in certain positions
to cells. By doing so, we are constructing a discrete 
``function'' $y = f (\vec{x})$, where 
$y$ runs in the ``attributes'' and $\vec{x} \in \{ M^{\otimes p} \}$
belongs to our $p$-dimensional space. 
We \underline{define} the phase space $\{ \Psi (N) \}$
as the space of the assignments, the
``maps'' $\Psi$:
\be
\Psi: \, N \, \to \, \prod_i \otimes M_i^{\otimes p_i} \, , ~~~~~ M_i 
\, \geq \, N \, . 
\label{psidef}
\ee 
For large $M_i$ and $N$, we can approximate
the discrete degrees of freedom with continuous coordinates: $M_i \to r_i$,
$N \to R$.
We have therefore a $p$-dimensional space with volume $\prod r_i^{p_i}$, and
a continuous map $y \in \{ R \} 
\stackrel{\Psi}{\to} \vec{x} \in \{ \vec{r}^{\vec p} \}$,
where $R \leq r_i$ $\forall i$. 
In the following we will always consider $M_i \gg N$, while keeping $V$
finite. This has to considered as a regularization condition,
to be eventually relaxed by letting $V \to \infty$.

The assignments \ref{psidef} are basically assignments of binary codes.
However, if we call $N$ the total energy, and the $M$ space coordinates,
it is clear that the $\Psi(N)$ are assignments of geometries, and that
$\{ \Psi (N) \}$ is the phase space of all the possible geometries
at energy $N$. To stay general, let us call them ``configurations''. 
In order to appropriately compare
configurations through the corresponding geometries, we may think 
of fixing the highest dimensionality of space, say $P$, fix a volume $V_P$ 
of this $P$-dimensional space~\footnote{Indeed, $P \leq V$ because it 
does not make sense
to speak of a space direction with less than one space cell.}, and work 
with the subclass of configurations that correspond to spaces of dimension
$p \leq P$, and volume smaller than $V_P$. In this way, all the geometries can
be thought as embedded in a common, higher space. $P$ and $V_P$ will then be 
let to go to infinity.

We want to investigate now what is the entropy of a certain
configuration in this phase space. 
An important observation is that
\emph{there do not exist two configurations with the same entropy}:
if they have the same entropy, they are perceived as the same configuration.
\label{noneq}
The reason is that we have a combinatoric problem, and,
at fixed $N$, the volume of occupation
in the phase space is related to the symmetry group of the configuration.
In practice, we classify configurations through combinatorics: 
a configuration corresponds to a certain combinatoric group. Now,
discrete groups with the same volume, i.e. the same number of elements,
are homeomorphic. This means that they describe
the same configuration. Configurations and entropies are therefore
in bijection with discrete groups, and this removes the degeneracy.
Different entropy = different occupation volume = different volume of the
symmetry group; in practice  this means that we have a different configuration.
The most entropic configurations are the 
``maximally symmetric'' ones, i.e.
those that look like spheres in the above sense.

\subsection{Entropy of spheres}
\label{eSp}

In order to compute the entropy of a sphere, we proceed as follows.
Let us consider distributing the $N$ energy attributes
along a $p$-sphere of radius $m$, $m \leq M$; 
our problem is then to find out
what are the most entropic ways of occupy $N$ of the $\sim m^p$ cells of the
sphere~\footnote{For simplicity we neglect numerical coefficients, because 
we are interested here in the scaling, for large $N$ and $m$.
This is also the reason why we use the terminology of the 
geometry on the continuum.}.
For any dimension, the most symmetric configuration is of course the one 
in which one
fulfils the volume, i.e. $N \sim m^p$. However, we are bound to the constraint
$N \leq m$ for any coordinate, otherwise we loose the interpretation 
at the ground of the whole construction, namely of $N$
as the coordinate of attributes, and $m$ as the target of the assignment. 
$N \sim m^p$ means $m \sim \sqrt[p]{N}$,
which implies $m < N$. The highest entropy we can attain is therefore
obtained with the largest possible value of $N$ as compared to $m$, i.e.
$N = m$, where once again the equality is intended up to an appropriate,
$p$-dependent coefficient. 
Let us start by considering the entropy of a three-sphere.
The weight in the phase space
will be given by the 
number of times such a sphere can be formed by moving along
the symmetries of its geometry, times
the number of choices of the position of, say, its centre, in the whole space.
Since we eventually are going to take the limit $V \to \infty$,
we don't consider here this second contribution, which
is going to produce an infinite factor, equal for each kind of geometry, for 
any finite amount of total energy $N$. We will therefore concentrate here
on the first contribution, the one that from three-sphere and other geometries.
To this purpose, we solve
the ``differential equation'' (more properly, a finite difference equation)
of the increase in the combinatoric when passing from $m$ to $m + 1$.
Owing to the multiplicative structure of the phase space 
(composition of probabilities),
expanding by one unit the radius, or equivalently the scale of all 
the coordinates, means
that we add to the possibilities to form the configuration for
any dimension of the sphere some more $\sim m+1$ times
(that we can also approximate with $m$, because we work at large $m$) 
the probability of one cell
times the weight of the configuration of the remaining $m$ 
(respectively $m-1$) cells.
But this is not all the story: since distributing $N$ energy cells along a 
volume
scaling as $\sim m^3$, $m \geq N$ means that our distribution does not
fulfill the space, the actual symmetry group of the distribution will be
a subgroup of the whole group of the pure "geometric" symmetry: moving
along this space by an amount of space shorter than the distance between
cells occupied by an energy unit will not be a symmetry, because one
moves to a "hole" of energy. It is easy to realize that
in such a "sparse" space, the effective symmetry group will have a 
volume that stays 
to the volume of a fulfilled space in the same ratio as the respective energy
densities. Taking into account all these effects, we obtain the following 
scaling:
\be
W(m+1)_3 \, \sim \, W(m)_3 \times   (m+1)^3 \times {N \over m^3} 
\times {m \over N} \, . 
\label{Wmm0}
\ee 
The last factor expresses the density of a circle,  
whereas the
factor ${N \over m^3}$ is the density of the three-sphere. 
In order to make the origin of the various terms more clear, in these 
expressions
we did not use explicitly the fact that actually $N$ is going
to be eventually identified with $m$. 
Indeed, in \ref{Wmm0} there should be one more factor:
when we pass from radius $m$ to $m+1$ while keeping $N$ fixed, 
the configuration becomes less dense, and we loose a symmetry factor
of the order of the ratio of the two densities:
$[m / (m+1 )] ^3 \sim 1 + {\cal O}(1 / m)$.
Expanding $W(m+1)$ on the left hand side
of \ref{Wmm0} as $W(m) \, + \, \Delta W(m)$, and neglecting on the r.h.s. 
corrections of order $1 / m$,
we can write it as:
\be
{\Delta W(m)_3 \over W(m)_3 } \, \simeq \, m \, . 
\label{dwwm3}
\ee
Since we are interested in the behaviour at large $m$, we can approximate it
with a continuous variable, $m \to x$, $x$, and approximate the finite
difference equation with a differential one. Upon integration, we obtain:
\be
S_{3} \, \propto \, \ln W(m)_3 \, \sim \, {1 \over 2} \, m^{2} \, ,
\label{Spm3}
\ee
where it is intended that $N = m$. Without this identification, the factor
$(m/ N)$ in \ref{Wmm0} would not be the density of a 1-sphere.
Under this condition, the energy density of the three-sphere
scales as $1 / N^2$,
and we obtain an equivalence between energy density and curvature $R$:
\be
\rho_3 (N) \, \sim \, {1 \over N^2} ~  \cong ~ {1 \over r^2} \, 
\sim \,  R_{(3)} \, .
\label{r3N}
\ee 
This is basically the Einstein's equation relating the curvature of space
to the tensor expressing the energy density. 
Indeed, here this relation can be assumed to be the physical description of a sphere
in three dimension. We can certainly think to formally distribute the $N$ energy units
along any kind of space with any kind of geometry, but what makes a curved
space \emph{physically} distinguishable from a flat one, and a particular geometry from another one? 
Geometries are characterized by the curvature, but how does one observer 
measure the curvature? The coordinates $m$
of the target space have no meaning without energy units distributed
along them. The geometry is decided by the way we assign the $N$
occupation positions. 
Here therefore we \emph{assume} that measuring
the curvature of space is nothing else than measuring the energy density.
For the time being, let us just take the equivalence between energy density and curvature as purely formal;
we will see in the next sections that this, with our definition of energy, will also imply 
that physical particles move along geodesics of the so characterized space, 
precisely as one expects from the Einstein's equations. 
We will come back to these
issues in section~\ref{relativity}.
In a generic dimension $p \geq 2$ the condition for having the geometry of
a sphere reads \footnote{We recall that we omit here $p$-dependent numerical coefficients
which characterize the specific normalization of the curvature of a sphere
in $p$ dimensions, because we are interested in the scaling at generic $N$, and $m$,
in particular in the scaling at large $N$.}:
\be
\rho_p (E) \, \sim \, {N \over m^p} ~ \cong ~ {1 \over m^2} \, . 
\label{rpN}
\ee
In dimension $p \geq 3$ it is solved by:
\be
m \, \sim \,  N^{1 \over p-2} ~ < \, N \, , ~~~ p \geq 3 \, .
\label{mphigher}
\ee
In two dimensions, \ref{rpN} implies $N = 1$ (up to some numerical coefficient). This means that, although it is technically possible to distribute $N > 1$
energy units along a two-sphere of radius $m > 1$, from a physical point of view
these configurations do not describe a sphere. This may sound strange, because
we can think about a huge number of spheric surfaces existing in our physical
world, and therefore we may have the impression that attempting to
give a characterization of the physical world in the way we are here doing
already fails in this simple case. The point is that all the two spheres of our
physical experience do not exist as two-dimensional spaces alone, but only as embedded
in a three-dimensional physical space. i.e. as subspaces of a three-dimensional
space.
In dimensions higher than three, the equivalent of \ref{Wmm0} reads:
\be
W(m+1)_p \, \sim \, W(m)_p \times   (m+1)^p \times {N \over m^p} 
\times {m \over N} \, . 
\label{Wm+1pN}
\ee
The last term on the r.h.s. is actually one, because it was only formally written
as $N / m$ to keep trace of the origin of the various terms. Indeed, it
indicates the density of a fulfilling space, to which the scaling of the weight of any
dimension must be normalized.
Inserting the condition for the $p$-sphere, equation~\ref{rpN}, we obtain:
\be
W(m+1)_p \, \sim \, W(m)_p \times   (m+1)^p \times {1 \over m^2} \, , 
\label{Wm+1p}
\ee
which leads to the following finite difference equation:
\be
{\Delta W (m)_p \over W(m)_p} ~ \approx ~
m^{p-2} \, .
\label{Wmp-2}
\ee
This expression obviously reduces to~\ref{dwwm3} for $p = 3$.
Proceeding as before, by transforming the finite difference equation
into a differential one, and integrating, we obtain:
\be
S_{( p \geq 2)} \, \propto \, \ln W(m) \, \sim \, {1 \over p-1} \, m^{p-1} \, , ~~~~
p \geq 3 \, .
\label{Spm}
\ee
This is the typical scaling law of the entropy of 
a $p$-dimensional black hole (see for instance \cite{Rabinowitz:2001ag}).
For $p=2$, if we start
from~\ref{Wm+1pN}, without imposing the condition~\ref{rpN}
of the sphere, we obtain, upon integration:
\be
S_{( 2)} \, \sim \, N^2 \, ,
\label{Sp2mN}
\ee 
formally equivalent to the entropy of a sphere in three dimensions.
However, the fact that the condition of the sphere \ref{rpN} implies
$N = 1$ means that a homogeneous distribution of the $N$ energy
units corresponds to a staple of $N$ two-spheres.
Indeed, if we use \ref{Wm+1p} and \ref{Wmp-2},
for which the condition $N = 1$ is intended, we obtain:
\be
S_{( 2)} \, \sim \, m \, .
\label{Sp2m}
\ee
For a radius $m = N$, this gives $1 / N$ of the result~\ref{Sp2mN},
confirming the interpretation of this space as the superposition of $N$
spheres. From a physical point of view, we have therefore $N$ times
the repetition of the same space, whose true entropy is not
$N^2$ but simply $N$. As we will see in the next sections,
such a kind of geometries correspond to what we will interpret as
quantum corrections to the geometry of the universe.
In the case of $p = 1$, from a purely formal point of view the condition
of the sphere~\ref{rpN} would imply $N = 1/m$. Inserted in~\ref{Wm+1pN}
and integrated as before, it gives:
\be
S_{( 1)} \, \propto \, \ln W(m) \, \sim \, \ln m \, , ~~~~~~
p = 1 \, .
\label{Sp1m}
\ee
Indeed, in the case of the one-sphere, i.e. the circle,
one does not speak of Riemann curvature, proportional to $1 / r^2$,
but simply of inverse of the radius of curvature, $1 / r$.
It is on the other hand clear that the most entropic
configuration of the one-dimensional space is obtained by a complete
fulfilling of space with energy units, $N=m$, and that the weight in the
phase space of this configuration is simply:
\be
W(N)_1 \, \sim \, N \, ,
\ee
in agreement with~\ref{Sp1m}~\footnote{We always factor out the group of permutations, which brings a volume factor $N!$ common to any
configuration of $N$ energy cells.}.
For the spheres in higher dimension, 
from expression~\ref{Spm} and~\ref{mphigher}we derive:
\be
S_{(p \geq 3)} \vert_N \, \sim \, {1 \over p-1} \, 
m^{p-1} \, \sim \, {1 \over p-1} \, N^{p-1 \over p -2} \, .
\ee 
For large $p$ the weights tend therefore to a $p$-independent value:
\be
W(N)_p ~ \stackrel{p \ggg 3}{\longrightarrow} ~ \approx \, {\rm e}^{N} \, , 
\label{Wlargep}
\ee
and their ratios tend to a constant. As a function
of $N$ they are exponentially suppressed
as compared to the three-dimensional sphere.
The scaling of the effective entropy as a function of $N$ 
allows us to conclude that:

$\bullet$ \emph{At any energy $N$, the 
most entropic configuration is the one corresponding to the geometry of a 
three-sphere. Its relative entropy scales as $S \sim N^2$}.

Spheres in different dimension have an unfavoured ratio
entropy/energy. Three dimensions are then statistically
``selected out'' as the dominant space dimensionality.

\subsection{The ``time'' ordering}
\label{timev}

A property of $\{ \Psi (N) \}$ is that, 
if $N_1 < N_2$ $\forall \Psi(N_1) \in \{ \Psi (N_1) \}$
$\exists \Psi^{\prime} (N_2) \in \{ \Psi (N_2) \}$ 
such that $\Psi^{\prime} (N_2)
\supsetneq \Psi (N_1) $, something that, with an abuse of language, we
write as:  
$\{ \Psi(N_2) \} \supset \{ \Psi(N_1) \}$, $\forall ~ N_1 < N_2$. 
It is therefore natural to
introduce an ordering in the whole phase space, that
we call a ``time-ordering'', through the identification 
of $N$ with the time coordinate: $N \leftrightarrow t$. 
We call ``history of the Universe'' the ``path'' $N \to \{ \Psi (N) \}$
\footnote{Notice that $\{ \Psi(N) \}$, the ``phase space at time $N$'', 
includes also tachyonic configurations.}. 
This ordering turns out to quite naturally correspond to our
everyday concept of time-ordering. 
In our normal experience, the reason why
we perceive a history basically consisting in a progress toward
increasing time lies on the fact that higher times bear the ``memory'' 
of the past, lower times.
The opposite is not true,  because ``future'' configurations are not contained
in those at lower, i.e. earlier, times. But in order to
be able to say that an event $B$ is the follow up of $A$, $A \neq B$ 
(time flow from $A \to B$), at the time we observe $B$ we need to
also know $A$. This precisely means $A \in \{ \Psi(N_A) \}$ \underline{and}
$A \in \{ \Psi(N_B) \}$, which implies $\{ \Psi(N_A) \} 
\subset \{ \Psi(N_B) \}$ in the
sense we specified above.
Time reversal is not a symmetry of the system~\footnote{Only by restricting to
some subsets of physical phenomena one can approximate the description 
with a model symmetric under reversal of the time coordinate, at the price
of neglecting what happens to the environment.}.

\subsection{How do inhomogeneities arise}
\label{inho}

In this set-up configurations are basically identified by their 
symmetry group. Configurations that describe the same geometry, but are 
``rotated'' with respect to each other, as compared to an external 
reference frame, actually describe \emph{the same} configuration.
The reason is that there is no ``external frame'': reference points are
defined through the intrinsic asymmetries of the configurations in themselves.
Reference points are introduced through asymmetries.
Starting from the most entropic one, we can think to progressively obtain
all the less entropic configurations 
by ``moving'' away the more and more units of energy
to form less and less symmetric configurations, also walking 
through different dimensions. 
In this way, one obtains a tower of 
asymmetric configurations ``stapled'' on the point at which the first 
asymmetry has been introduced. 
The superposition of configurations does not produce therefore
a uniform universe, but a kind of ``spontaneous'' breaking of any symmetry.
From the property, stated at page~\pageref{noneq}, 
that at any time ${\cal T} \sim N$
there do not exist two inequivalent
configurations with the same entropy, and from the fact that less
entropic configurations possess a lower degree of symmetry, we obtain
that:

$\bullet$ 
\emph{At any time ${\cal T}$ the average appearance of the universe is that
of a space in which \bf{all symmetries are broken}.}  

The amount of breaking, depending on the weight of non-symmetric
configurations as compared to the maximally symmetric one, involves
a relation between the energy 
(i.e. the deformations of the geometry) and the time
spread/space length, of the deformation, that will be
discussed in the next sections. 
The inhomogeneities produced in this way give rise to the varied spectrum of
mass and energy clusters, galaxies, particles, etc...
As there is no external frame,
in this framework there is also no external observer: 
an observer is a ``local inhomogeneity'' of space, and necessarily
belongs to the universe. 
The observer is only sensitive to 
its own configuration, in the sense that he ``learns'' 
about the full space only through
the superposition of configurations he is made of, and their changes. 
For instance, he can perceive that the configurations
of space of which he is built up change with time, and \emph{interprets}
these changes as due to the interaction with an environment.

\subsection{Mean values and observables}
\label{vev}

At any time ${\cal T} \sim N$ in the ``universe'' given by $\{ \Psi (N) \}$
the mean value of any observable quantity ${\cal O}$ is
the sum of the contributions to ${\cal O}$
over all configurations $\Psi$, weighted according to their volume of 
occupation in the phase space:
\be
< {\cal O} > \; \propto \, \sum_{\Psi ({\cal T})} 
W (\Psi)\, {\cal O}(\Psi) 
\, .
\ee
We have written the symbol $\propto$ instead of $=$ because,
as it is, the sum on the r.h.s. is not normalized. 
The weights don't sum up to 1, and not even do they sum up to a finite
number: in the infinite volume limit, they all diverge
\footnote{As long as the volume, i.e. the total number of cells of
the target space, for any dimension, is finite, there is only a finite 
number of ways one can distribute energy units. 
In the infinite volume limit, both the number of possibilities
for the assignment of energy, and the
number of possible dimensions, become infinite.}. However, as we 
discussed in section~\ref{ddf}, what matters is their 
relative ratio, which is finite because the infinite volume factor
is factored out. In order to normalize mean values,
we introduce a functional that works as ``partition
function'', or ``generating function'' of the Universe:
\be
{\cal Z} \, \stackrel{\rm def}{=} \sum_{\Psi ({\cal T})} 
W(\psi)   
\, = \, \sum_{\Psi ({\cal T})} 
{\rm e}^{S(\Psi)} \, .
\label{ZPsi}
\ee
The sum has to be intended as always performed at finite volume.
In order to define mean values and observables,
we must in fact always think in terms of finite space volume, 
a regularization condition to be eventually relaxed. 
The mean value of an observable can then be written as:
\be
< {\cal O} > \, \stackrel{\rm def}{\equiv} \, {1 \over {\cal Z}}
\sum_{\Psi ({\cal T})} W (\Psi)\, {\cal O}(\Psi) \, .
\label{meanO}
\ee
Mean values therefore are not defined in an absolute way, 
but through an averaging procedure in which
the weight is normalized to the total weight of all the configurations,
at any finite space volume $V$.

\subsection{Summing up geometries}
\label{sumgeo}

We may now ask what a ``universe'' given by the collection of
all configurations at a given time $N$ looks like to an observer.
Indeed, a physical observer will be part of the universe, and as such
correspond to a set of configurations that identify a preferred point, 
something less symmetric and homogeneous than a sphere. 
However, let us just assume that the observer
looks at the universe from the point of view of the most entropic 
configuration, namely it lives in three dimensions, and interprets 
the contribution of any configuration in terms
of three dimensions. This means that he will not perceive the universe
as a superposition of spaces with different dimensionality, but will
measure quantities, such as for instance energy densities,
referring them to properties of the three dimensional space, although
the contribution to the amount of energy may come also
from configurations of different dimension (higher or lower than three).

From this point of view, let us see how the contribution  
to the average energy density of space
of all configurations which are not the three-sphere is perceived. 
In other words, we must see how do the $p \neq 3$ configurations
project onto three dimensions.
The average density should be given by:
\be
\langle \rho(E)  \rangle ~ = ~
{\sum_{\Psi(N)} W(\Psi(N)) \rho(E)_{\Psi (N)} 
\over
\sum_{\Psi(N)} W(\Psi(N)) } \, .
\ee
We will first consider the contribution of spheres.
To the purpose, it is useful to keep in mind that at fixed $N$ 
(i.e. fixed time) higher
dimensional spheres become the more and more ``concentrated'' around
the (higher-dimensional) origin, and the weights tend
to a $p$-independent value for large $p$ (see \ref{mphigher} 
and \ref{Wlargep}).
When referred to three dimensions,
the energy density of a $p$ sphere, $p > 3$, is
$1 / N^{p-1}$, so that, when integrated over the volume, which
scales as $\sim N^p$, it gives a total energy $\sim N$.
There is however an extra factor $N^3 / N^{p} $ due to the fact that
we have to re-normalize volumes to spread all the higher-dimensional energy
distribution along a three-dimensional space.
All in all, this gives a factor $1 / N^{2(p-2)}$ in front of
the intrinsic weight of the $p$-spheres. Since the latter depend
in a complicated exponential form on $P$ and $N$, it is not possible
to obtain an expression of the mean value of the energy distribution
in closed form. However, as long as we are interested in just giving 
an approximate estimate, we can make several simplifications.
A first thing to consider is that, as we already remarked, at finite
$N$, the number of possible dimensions is finite, because
it does not make sense to distribute less than one 
unit of energy along a dimension: as a matter of fact such a space would
not possess this dimension. Therefore, $p \leq N$.
In the physically relevant cases $N \ggg 1$, and
we have anyway a sum over a huge number of terms, so that we can approximate
all the weights but the three dimensional one by their asymptotic value,
$W \sim \exp N$. This considerably simplifies our computation, because
with these approximations we have:
\be
\langle \rho(E)_N  \rangle ~ \approx ~
{1 \over {\rm e}^{N^2} \, + \, N \, {\rm e}^N} \times
\left[ {1 \over N^2} \, {\rm e}^{N^2} \, + \, \sum_{p > 3} 
{1 \over N^{2(p-2)}} \, {\rm e}^{N} \right] \, ,
\ee 
that, in the further approximation that $\exp N^2 \gg N \exp N$,
so that $\exp N^2 \, + \,  N \exp N \, \approx \, \exp N^2$,
we can write as:
\be
\langle \rho(E)_N  \rangle ~ \approx ~
{1 \over N^2} ~ + ~ {\rm e}^{- N} \, \left[ {1 \over 1 - {1 \over N^2}}
\right] ~ \approx ~ {1 \over N^2} ~ + ~ {\cal O} \left( {\rm e}^{- N} \right)
\, .
\label{rhoE3}
\ee 
We consider now the contribution
of configurations different from the spheres.
Let us first concentrate on the dimension $D = 3$, which is the most relevant 
one.
The simplest deformation of a 3-sphere consists in moving just one energy unit
one step away from its position on the sphere. Owing to this move, we 
break part of the symmetry. Further breaking is produced by moving more
units of energy, and by larger displacements. 
Our problem is to estimate the amount of reduction of the weight as compared
to the sphere. Let us consider displacing just one unit of energy.
We can consider that the overall symmetry group of the sphere is
so distributed that the local contribution is proportional to the density 
of the sphere, $1 / N^2$. Displacing one unit energy cell should then reduce 
the overall weight by a factor $\sim (1 - {1 / N^2})$. Displacing the unit 
by two steps 
would lead to a further suppression of order
$1 / N^2$. Displacing more units may lead to partial symmetry restoration
among the displaced cells. Even in the presence of
partial symmetry restorations the suppression factor due to the
displacement of $n$ units remains of order
$\approx n^2 / N^{2n}$ (the suppression factor divided
by the density of a sphere made of $n$ units) as long as $n \ll N$. 
The maximal effective
value $n$ can attain in the presence of maximal
symmetry among the displaced points is of course $N/2$, beyond
which we fall onto already considered configurations. 
This means that summing up all the contributions
leads to a correction which is of the order of the sum of an (almost) 
geometric series of ratio $1 / N^2$. 
Similar arguments can be applied to $D \neq 3$, to conclude that
expression \ref{rhoE3} receives all in all a correction of order $1 / N^2$.
This result is remarkable. As we will discuss in the following along this
paper, the main contribution to the geometry of the universe, the one
given by the most entropic configuration, can be viewed as the
classical, purely geometrical contribution, whereas those given by 
the other, less entropic geometries, can be considered contributions to the
"quantum geometry" of the universe~\footnote{In Ref.~\cite{npstrings-2011}
we discuss how the classical part of the curvature can be
referred to the cosmological constant, while the other terms to
the contribution due to matter and radiation. In particular, we recover
the basic equivalence of the order of magnitude of these contributions, as
the consequence of a non-completely broken
symmetry of the quantum theory which is going to represent
our combinatorial construction in terms of quantum fields and particles.}. 
From \ref{rhoE3} we see that not only the three-dimensional
term dominates over all other ones, but that it is
reasonable to assume that 
\emph{the universe looks mostly like three-dimensional}, indeed
mostly like a three-sphere.
This property becomes stronger and stronger as time goes by 
(increasing $N$).
From the fact that the maximal entropy is the one of
three spheres, and scales as $S_{(3)} \sim N^2$, we derive also that
the ratio of the overall weight of the configurations at time $N-1$,
normalized to the weight at time $N$, is of the order:
\be
W(N-1) \, \approx \, W(N) \, {\rm e}^{-2N} \, .
\ee
At any time, the contribution of past times is therefore negligible as compared
to the one of the configurations at the actual time. The suppression factor
is such that the entire set of three-spheres at past times sums up 
to a weight of the order of $W(N-1)$:
\be
\sum_{n =1}^{N-1} W(n) ~ \approx ~ \sum {1 \over ({\rm e}^2)^n} ~ \sim ~
{\cal O}(1) \, .
\label{sumnN}
\ee
We want to estimate now the overall contribution to the partition function
due to all the configurations, as
compared to the one of the configuration of maximal entropy.
We can view the whole spectrum of configurations as obtained by
moving energy units, and thereby deforming parts of the symmetry, starting
from the most symmetric (and entropic) configuration. In this way,
not only we cover all possible configurations in three dimensions, but we
can also walk through dimensions. 
In order to account for the contribution
to the partition function of all
the deformations of the most entropic
geometry, we can think of a series of steps, in which we move from
the spheric geometry one, two, three, and so on, units of symmetry.
At large $N$, we can approximate sums with integrals, and
account for the contribution to 
the ``partition function'' \ref {ZPsi} of all the configurations
by integrating 
over all the possible values of entropy, decreasing from the maximal one. 
In the approximation of variables on the continuum, symmetry groups
are promoted to Lie groups, and moving positions
and degrees of freedom is a ``point-wise'' operation that can be viewed
as taking place on the algebra, not on the group elements. Therefore,
the measure of the integral is such that we sum over incremental
steps on the exponent, that is on the logarithm of the weight, the entropy.
Therefore, we can write the sum over weights as a sum 
over the decrements from the highest entropy, the one of the 
three-sphere, given in~\ref{Spm}, namely $S_{\rm max} = S_0 =
\exp N^2$:
\be
{\cal Z} ~~ \gsim ~~  
\int^{S_0}_0 d L \; 
{\rm e}^{S_0 \left(1 - L \right) } \, .
\label{Zall1}
\ee
This has to be taken as 
an approximate way of accounting for the order
of magnitude of the contribution of the infinity of configurations.
Integrating~\ref{Zall1}, we obtain:
\be
{\cal Z} ~~ \approx ~~ {\rm e}^{S_0} \left(1 \, + \, {1 \over S_0}  \right)
\, .
\label{Zallint1}
\ee  
The result would however not change if, instead of considering 
the integration on
just one degree of freedom, parametrized by one coordinate, $L$, we would
integrate over a huge (infinite) number of variables, each one contributing
independently to the reduction of entropy, as in:
\be
{\cal Z} ~~ \approx ~~ \sum^N_{n=1} 
\int d^n L \; 
{\rm e}^{S_0 \left[1 -  (L_1 + \ldots + L_n) \right]} \, ,
\label{Zall}
\ee
In the second case, \ref{Zall}, we would have:  
\be
{\cal Z} ~~ \approx ~~ {\rm e}^{S_0} \sum_n {1 \over S_0^n } \, = 
\, {\rm e}^{S_0} \left( 1 + {1 \over S_0 -1} \right) \, ,
\label{Zallint}
\ee  
anyway of the same order as~\ref{Zallint1}. Together with \ref{sumnN},
this tells us also that instead of \ref{zsum1} we could as well define the
partition function of the universe at "time" ${\cal E}$ by the sum over all
the configurations at past time/energy $E$ up to ${\cal E}$:
\be
{\cal Z}_{\cal E} ~ = ~ \sum_{\psi(E \leq {\cal E})} {\rm e}^{S (\psi)} \, .
\label{zsum}
\ee

\section{The Uncertainty Principle}
\label{UncP}

According to \ref{meanO}, quantities which are observable by an
observer living in three dimensions do not receive contribution
only from the configurations of extremal or near to extremal entropy: 
all the possible configurations at a certain time
contribute. Their value bears therefore a ``built-in'' uncertainty,
due to the fact that, beyond a certain approximation, 
experiments in themselves cannot be defined as physical quantities of a 
three-dimensional world.
In section~\ref{timev} we have established the correspondence 
between the ``energy'' $N$ and the ``time'' coordinate that orders the history
of our ``universe''. Although we gave $N$ the interpretation of 
total energy of a configuration, and as such it determines its geometry,
this does not mean that $N$ is also the total energy of the universe, as it is
measured by an observer, necessarily belonging to the universe.
In other words, $N$ does not coincide with the
\emph{operational} way we define energy, related to the way we measure it.
Indeed, as it is, $N$ simply reflects the ``time'' coordinate, and states
the age of the universe. The energy one can measure is an average quantity
defined as in \ref{meanO}. In principle, since $E =N$ for any configuration, 
the two quantities seem to coincide, but as a matter of fact they don't:
re-summing $E = N$ from \ref{meanO} implies a knowledge of all the 
configurations contained in $\{ \Psi (N) \}$. 
The universe 
is the result of a superposition in which also very singular configurations 
contribute, in general uninterpretable within the usual conceptual framework 
of particles, or wave-packets, and in general of geometries
of a three-dimensional space. Therefore, not directly accessible to 
a three-dimensional observer.    
When we measure an energy, or equivalently a ``geometric curvature'', 
we refer therefore
to an average and approximated concept, for which we consider only a subset of
all the configurations of the universe. Now, we have seen that the 
larger is the ``time'' $N$, the higher is the dominance of the most probable
configuration over the other ones, and therefore more picked is the average,
the ``mean value'' of geometry. The error in the evaluation of the
energy content will therefore be the more reduced, 
the larger is the time spread one considers, 
because relatively lower becomes the weight of the configurations one ignores.
From \ref{Zallint} we can have an idea of what is the order of the 
uncertainty in the evaluation of energy.
According to \ref{Zall}
and \ref{Zallint}, the mean value of the total energy, receiving contribution 
also from all the other configurations, results to be ``smeared'' by an amount:
\be
< E  > ~ \approx ~ E_{S_0} \, + \, E_{S_0} \times \, {\cal O} 
\left( 1 / S_0  \right) \, .
\label{Emean} 
\ee  
That means, inserting $S_0 \approx N^2 \equiv t^2 \sim \, E^2_{S_0}$:
\be
< E  > ~ \approx ~ E_{S_0} \, + \, \Delta E_{S_0} ~ \approx ~ 
E_{S_0} \, + \, {\cal O} \left( {1 \over t} \right) 
\, .
\label{deltaEmean} 
\ee  
Consider a subregion of the universe, of extension 
$\Delta t$~~\footnote{We didn't yet introduce units distinguishing between 
space and time. In the usual language we could consider this region as being
of ``light-extension'' $\Delta x = c \Delta t$.}. 
Whatever exists in it, namely,
whatever differentiates this region from the uniform spherical ground geometry
of the universe, must correspond to a superposition of 
configurations of non-maximal entropy. 
From our considerations of above, we can derive that it is not possible to
know the energy of this subregion with an uncertainty lower than the inverse
of its extension. In order to see this, let's 
estimate what is the amount of the contribution to this energy given by the
sea of configurations of non-maximal entropy. As discussed,
these include higher and lower space dimensionalities, and any other kind
of differently interpretable combinatorics. 
The mean energy will be given as in \ref{Emean}.
However, this time the maximal entropy
$\tilde{S}_0(\Delta t)$ of this subsystem will be lower than the upper bound
constituted by the maximal possible entropy of a region enclosed in a time 
$\Delta t$, namely the one of a three-sphere of radius $\Delta t$:
\be
\tilde{S}_0(\Delta t) ~ < ~ \left[ \Delta t \right]^2 \, ,
\label{Stilde}
\ee 
and the correction corresponding to the
second term in the r.h.s. of \ref{deltaEmean} will just constitute a lower 
bound to the energy uncertainty \footnote{The maximal energy can be 
$E \sim \Delta t$ even for a class of non-maximal-entropy, non-spheric 
configurations.}:
\be
\Delta E ~~ \gsim ~~ {{\Delta t} \over S_0 (\Delta t)} \, ~~
\approx \, {1 \over \Delta t } \, . 
\label{Etuncertainty}
\ee
In other words, \emph{no region of extension $\Delta t$
can be said with certainty to possess an energy lower than $1/ \Delta t$}.
When we say that we have measured
a mass/energy of a particle, we mean that we have measured 
an average fluctuation of the configuration of the universe around the 
observer, during a certain time interval. 
This measurement is basically a process that takes place along the time 
coordinate. 
During the time of the ``experiment'', 
$\Delta t$, a small ``universe'' of superposing configurations
opens up for this particle. Namely, what we 
are probing are the configurations of a space region created in a time 
$\Delta t$, in which the highest entropy is the one of a sphere, this
time of radius $\Delta N = \Delta t$, etc... 
According to \ref{Etuncertainty}, the particle
possesses therefore a ``ground''
indeterminacy in its energy:
\be
\Delta E \, \Delta t ~\gsim ~ 1\, . 
\label{HUP}
\ee
As a bound, this looks quite like the time-energy Heisenberg
uncertainty relation~\footnote{Introducing the Planck constant
is here just a matter of introducing units enabling to measure energies 
in terms of time.}. 
In the case we consider the whole universe itself,
expression \ref{Zallint}
tells us that the terms neglected in the partition function,
due to our ignorance of the ``sea'' of all the possible configurations
at any fixed time, contribute to an ``uncertainty''
in the total energy of the same order as the inverse of the age of the 
universe:
\be
\Delta E_{\rm tot} \, \sim \, {\cal O} \left( 1 \over {\cal T}  \right)
\, .
\label{DeL}
\ee
Namely, an uncertainty of the same order as the imprecision due
to the bound on the size of the minimal energy steps at time ${\cal T}$. 
The quantity
$1/ S_0 \sim 1 / {\cal T}^2$ basically corresponds to 
the parameter usually called ``cosmological constant'', 
that in this scenario is not constant~\footnote{The approximate
value of the cosmological constant is usually computed to be
$\Lambda \sim 1 / H^2_0$, where $H_0$ is the Hubble parameter, whose value
corresponds, under appropriate conversion of units, to the present age of the 
universe.}. 
The cosmological constant therefore 
corresponds to a bound on the effective precision of calculation
of the predictions of this theoretical scenario.
\ref{HUP} and \ref{DeL} tell us that 
theoretical and experimental uncertainties are of the same order.
The bound to an experimental access to the
universe as we know it corresponds to the limit within which such a
universe is in itself defined. Beyond this threshold, there is a ``sea''
of configurations in which  
i) the dimensionality of space is not fixed;
ii) interactions are not defined, iii) there are tachyonic contributions,
causality does not exist etc...
beyond this threshold there is a sea of...uninterpretable 
combinatorics.

$\bullet$ \emph{It is not possible to go beyond the Uncertainty 
Principle's bound with the precision in the measurements, because this bound 
corresponds to the precision with which the quantities to be measured 
themselves are defined}.

\section{Deterministic or probabilistic physics?}
\label{detprob}

The sum \ref{zsum1} implies a scenario which is neither probabilistic 
in the usual sense of quantum mechanics, nor deterministic according to 
the usual meaning of causality. 
It is rather "determined" by the partition function 
at any time.
The universe at time
$N^{\prime} \sim {\cal T}^{\prime} = {\cal T} + \delta {\cal T} \sim N +1$ 
is not obtained by running forward, possibly through 
equations of motion, the configurations at time 
$N \sim {\cal T}$, it is not their 
``continuation'': it is given
by the weighted sum of all the configurations at time
${\cal T} + \delta {\cal T}$, as the universe at time ${\cal T}$ was
given by the weighted sum of all the configurations at time ${\cal T}$. 
In the large $N$ limit,
we can speak of ``continuous time evolution'' only 
in the sense that for a small change of time, the dominant
configurations correspond to distributions
of geometries that don't differ that much from those at previous time.
With a certain approximation we can therefore speak of evolution in 
the ordinary sense of (differential, or difference) time equations. 
Owing to the fact that at any time the appearance of the universe 
is mostly determined by the most entropic configurations, in the average

$\bullet$ {\emph{the dynamics of the evolution of the system is of entropic 
type.}}

\noindent
On the other hand, a full knowledge of the infinite terms of \ref{zsum1}
is impossible, and, owing to the fact that configurations in any
dimensions are accounted, also ill-defined. From this point
of view, the probabilistic interpretation
of the Heisenberg Uncertainty given in quantum mechanics seems
a viable way of parametrizing the unknown, reintroducing thereby a certain
degree of predictability and calculability. This is also the case of
systems in which the asymmetries are "hidden" below the threshold of the 
uncertainty \ref{HUP}, and produce therefore 
the impression of
equal probability of equivalent situations, 
like the two possible paths of an electron in the double slit experiment: 
being able to predict the details of an event, such
as for instance the precise position each electron will hit on the plate,
and in which sequence, requires to know the function
``entropy'' for an infinite number of configurations,
corresponding to any space dimensionality at fixed ${\cal T} \approx N$,
for any time ${\cal T}$ the experiment runs on.
Clearly, no computer or human being can do that. 
If on the other hand we content ourselves with an approximate predictive 
power, we can roughly reduce physical situations to certain ideal schemes, 
such as for instance ``the symmetric double slit'' problem. 
Of course, from a theoretical point of view we lose  
the possibility of predicting the position 
the first electron will hit the target (something anyway practically 
impossible to do), but we gain, at the
price of introducing symmetries and therefore also concepts like
``probability amplitudes'', the capability of predicting with a good
degree of precision the shape an entire beam of electrons will draw on the
plate. We give up with the ``shortest scale'', and we concern ourselves
only with an ``intermediate scale'', larger than the point-like one,
shorter than the full history of the universe itself. 
The interference pattern arises as the dominant mean configuration,
as seen through the rough lens of this ``intermediate'' scale. 
In this scenario, quantum de-coherence is ``built-in'' in \ref{zsum1}.

\section{Relativity}
\label{relativity}

As we discussed in sections~\ref{eSp}--\ref{sumgeo}, 
although the volume of the target spaces of the maps $\Psi (N)$
is eventually to be considered infinite, $V \to \infty$,
at any finite time the dominant configuration of the universe 
corresponds to a three-sphere 
of radius $N \sim {\cal T}$. 
On the top of this staple many ``almost spherical'', three-dimensional
configurations that, in the superposition, give rise to a space
with energy clusters. But in the sum \ref{ZPsi} there are also 
configurations which correspond to a geometry
not bounded within a region of radius 
$N \sim {\cal T}$, nor three-dimensional. Indeed, for any $V$, 
there are configurations which ``fulfill'' the volume. They
contribute in the form of quantum perturbations, 
all of them falling under the ``cover'' of the Uncertainty Principle, and 
being therefore
related to what we interpret as the quantum nature of physical phenomena.
This can be interpreted in the following way: at any finite time
${\cal T}$ we have a universe which is infinitely extended, but 
that can be organized by separating it into a ``classical part'', with a geometry
looking like the interior of a black hole, with a horizon placed at distance
$\propto {\cal T}$, and a quantum part, which accounts for the contribution
of any other kind of configurations. Only the classical part
can be reduced to the ordinary 
geometric interpretation of space extended only up to a distance 
$\propto {\cal T}$. In this perspective,
\
\\

$\bullet$ \emph{the space ``outside'' the horizon is infinitely extended, 
but it contributes to the perception of a classical observer and to
the values of the observables defined in the three-dimensional classical 
space only through the uncertainty of mean values, accounted for by the
Heisenberg's uncertainty}.
\
\\

\noindent 
In the following we want to see how in this universe
Einstein's special (and general) relativity are implied as a particular limit,
in which one considers just the classical part of space.

\subsection{From the speed of expansion of the universe to a maximal speed for 
the propagation of information}
\label{speedlight}

\begin{figure}[ht]
\begin{minipage}[b]{0.31\linewidth}
\centering
\epsfxsize=5cm
\epsfbox{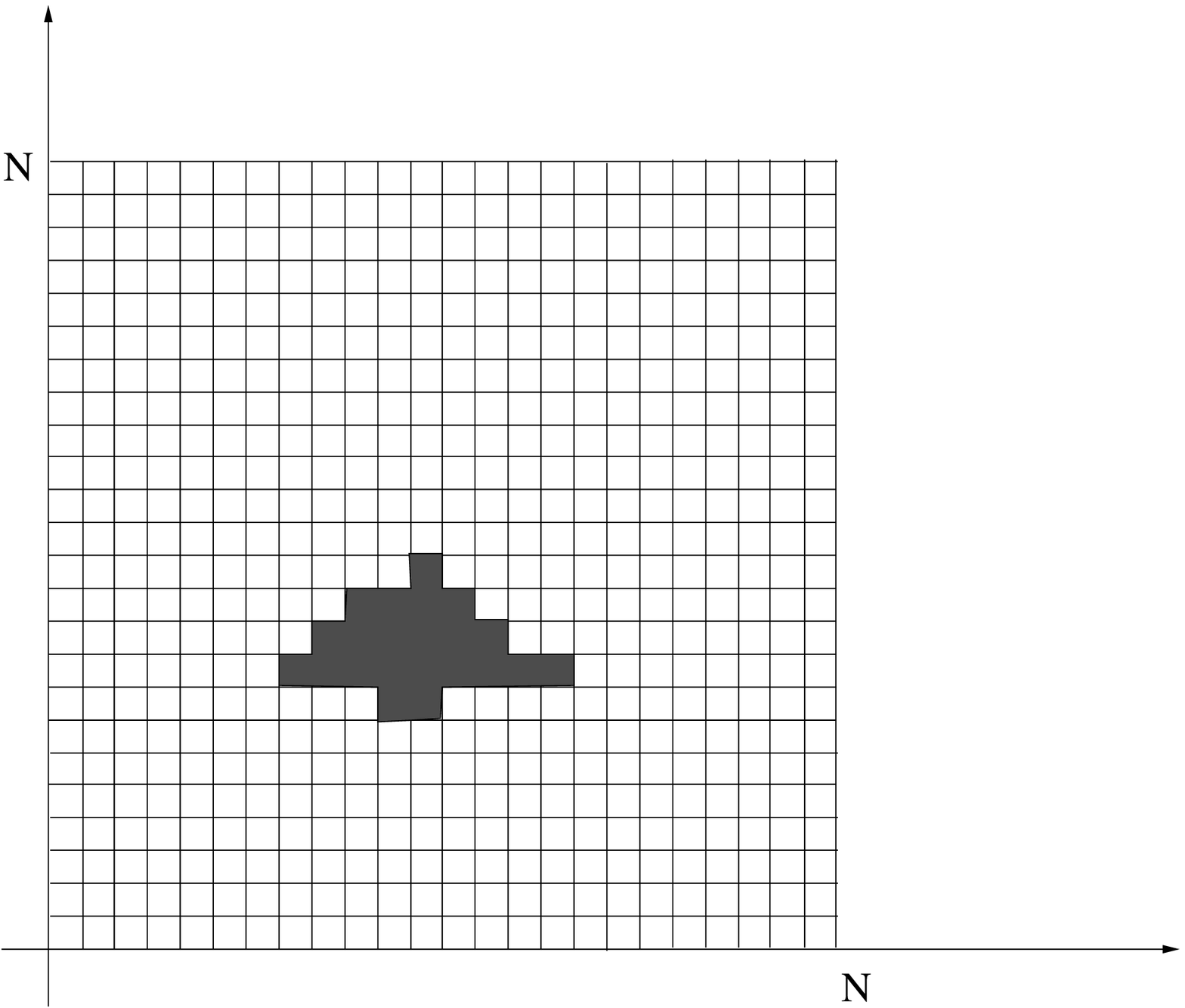}
\caption{}
\label{gridN}
\end{minipage}
\hspace{0.1cm}
\begin{minipage}[b]{0.31\linewidth}
\centering
\epsfxsize=5cm
\epsfbox{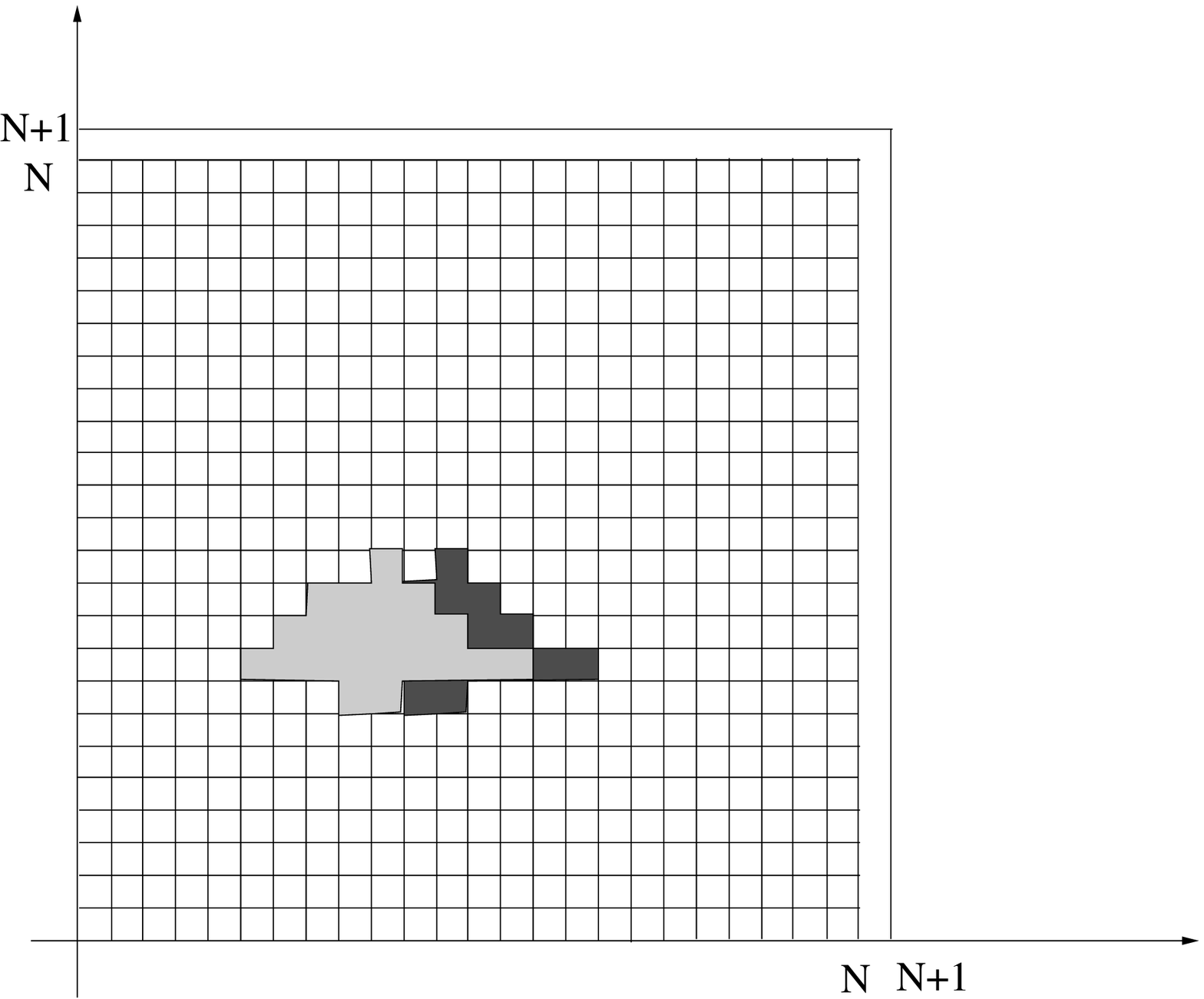}
\caption{}
\label{gridN+1}
\end{minipage}
\hspace{0.1cm}
\begin{minipage}[b]{0.31\linewidth}
\centering
\epsfxsize=5cm
\epsfbox{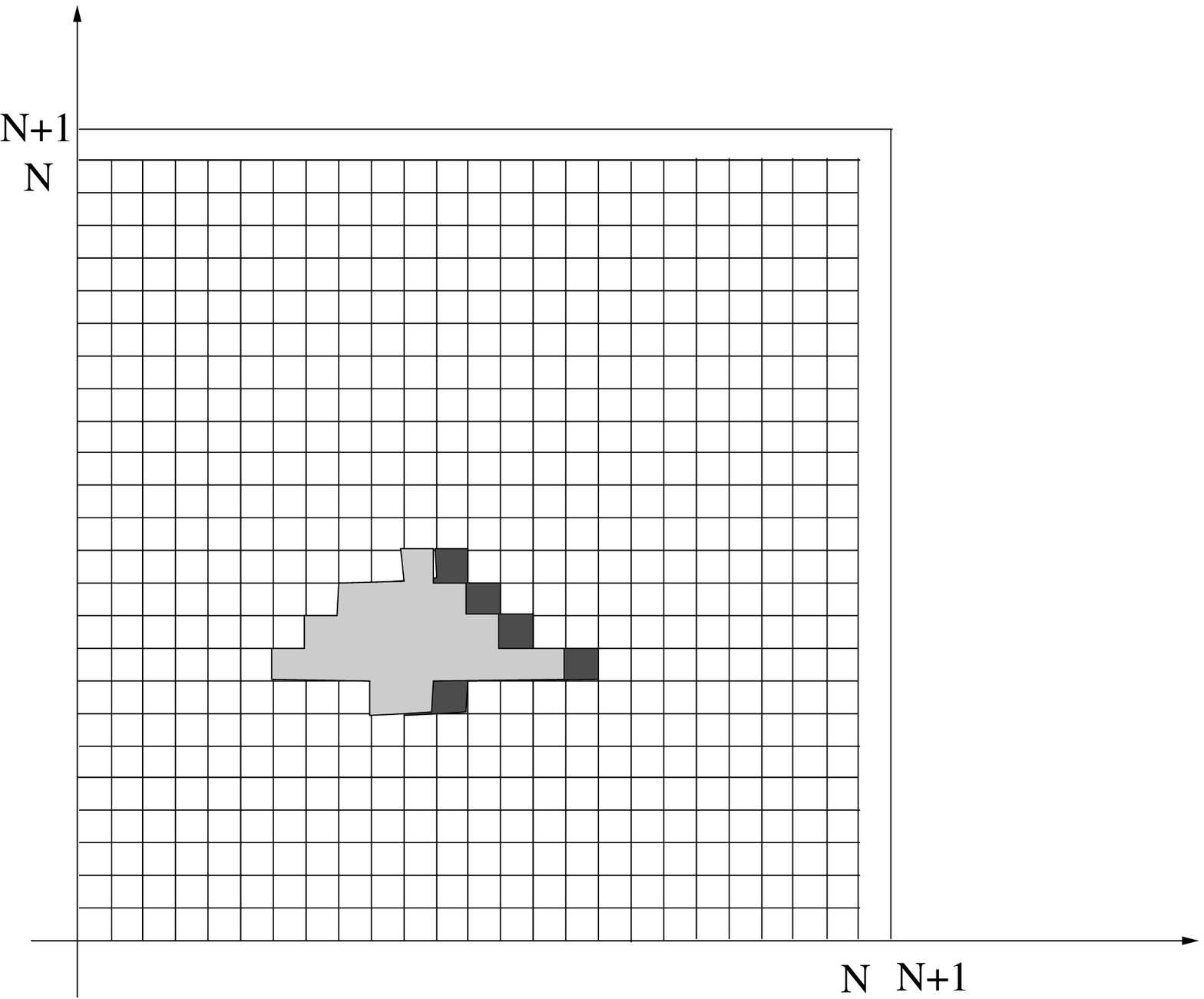}
\caption{}
\label{gridN+1bis}
\end{minipage}
\end{figure}

The classical space corresponds to a universe of radius $\sim N$ at time $N$, 
with total energy
also $N$. It expands at speed $1$. Indeed, we can introduce
a factor of conversion from time to space, $c$, and say that, by choice
of units, we set the speed of expansion to be $c = 1$ (in an obvious way, also
the conversion between units of space, and time, on one side, 
and energy on the other
side, is here ``by default'' set to one, but it can be called $h$).  
We want to see how this is also the 
maximal speed for the propagation of information within the classical space.
It is important to stress that all this refers only to the classical space
as we have defined it, 
because only in this sense we can say that the universe is three dimensional:
the sum \ref{ZPsi} contains in fact also configurations that,
through the time flow, can be interpreted as ``tachyonic'', along with
configurations in which it is not even clear what is the meaning of
speed of propagating information in itself, as there is no recognizable 
information at all, at least in the sense we usually intend it.
Indeed, when we say we get information about, say, the motion of a particle,
or a photon,
we intend to speak of a non-dispersive wave packet, so that we can say
we observe a particle, or photon, that remains particle, or photon,
along its motion \footnote{Like a particle, also a physical photon, 
or any other field,
is not a pure plane wave but something localized, therefore a 
superposition of waves, a wave packet.}. The existence,
in the scenario implied by \ref{ZPsi}, of structures of this 
kind, namely of wave packets that behave like massive particles,
or massless photons etc., is confirmed by the analysis
performed in \cite{npstrings-2011}.
Let's consider the simplified case of a universe
at time $N$ containing only one such a wave packet \footnote{We may think to
concentrate onto only a portion of the universe, where only such a wave
packet is present.}, as illustrated in 
figure~\ref{gridN}, where it is represented by the shadowed cells, and
the space is reduced to two dimensions.
Consider now the evolution at the subsequent instant of time,
namely after having progressed by a unit of time. 
Adding one point, $N \to N+1$, does 
produce an average geometry of a three 
sphere of radius $N+1$ instead of $N$. In the average, it is therefore like 
having added $4 \pi N^2$ ``points'', or unit cells. 
Remember that we work always with an 
infinite number of cells in an unspecified number of dimensions; when we talk 
of universe in three dimensions within a region of a certain radius, 
we just talk of the dominant geometry.
Let's suppose the position of the
wave packet jumps by steps (two cells) back, 
as illustrated in figure~\ref{gridN+1}. Namely, as time, and consequently also
the radius of the universe, progresses by one
unit, the packet moves at higher speed, jumping by two units. 
Compare this case with the case in which the packet jumps by just one unit,
as in figure~\ref{gridN+1bis}. 
The entropy of this latter configuration, intermediate between the first and 
second one, cannot be very different from the one of the second configuration,
figure~\ref{gridN+1}, in which the packet
jumps by two steps, because that was supposed to be the dominant configuration
at time $N+1$, and therefore the one of maximal entropy. Indeed, 
by ``continuity'' it must interpolate between step 2 and the configuration
at time $N$, that was also supposed to be a configuration of maximal
entropy. Therefore, the actual
appearance of the universe at time $N+1$ must be somehow a superposition
of the configurations 2 and 3, thereby contradicting our hypothesis that
the wave packet is non-dispersive~\footnote{If it was dispersive,
it would be something like a particle that, during its motion, ``dissolves'', 
and therefore we cannot anymore trace as a particle. It would be just
a ``vacuum fluctuation'' without true motion, something that does not carry
any information in the classical sense.}. 
Therefore, the wave packet cannot jump by two steps, 
and we conclude that the maximal speed allowed is that of expansion of the 
radius of the universe itself, namely, $c$.

It is too early here to discuss the actual existence in this scenario
of degrees of freedom that can be interpreted as photons.
In order to do this we must pass to a representation on 
the continuum, where, as discussed in~\cite{assiom-2011}, it corresponds to
a string scenario. Here we just anticipate that,
according to this theoretical framework,  
the reason why we have a universal bound on the speed of light
is therefore that light carries what we call classical information. 
Information about whatever kind of event tells 
about a change of average entropy of the observed system, 
of the observer, and what surrounds and connects them too. 
The rate of transfer/propagation
of information is therefore strictly related to the rate of variation of
entropy. Variation of entropy is what gives the measure of time progress
in the universe. Any carrier of information that ``jumps'' steps of the
evolution of the universe, going faster than its rate of entropy variation, 
becomes therefore dispersive, looses information
during its propagation. Light must therefore propagate at most at the rate of
expansion of space-time (i.e. of the universe itself). Namely, at the rate of  
the space/time conversion, $c$.

\subsection{The Lorentz boost}
\label{Lboost}

Let's now consider physical systems that can be identified 
as ``massive particles'', i.e. let us assume that there are local
superpositions of configurations which are interpreted as 
travelling at speeds always lower than $c$. 
Since the phase space has a multiplicative structure, 
and entropy is the logarithm of 
the volume of occupation in this space, it is possible to separate for each 
such a system the entropy into the sum of an internal, ``rest'' entropy, and
an external, ``kinetic'' entropy. The first one refers to the structure
of the system in itself, that can be a particle or an entire
laboratory. A point-like particle is an extended object of which
we neglect the geometric structure. 
The second one refers to the relation/interaction of this
system with the environment, the external world: its motion, the
accelerations and external forces it experiences, etc.      

\begin{figure}
\centerline{
\epsfxsize=4cm
\epsfbox{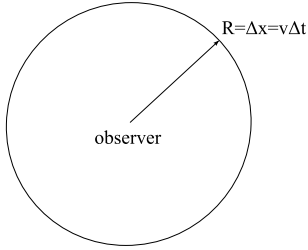}
}
\vspace{0.3cm}
\caption{During a time $\Delta t$, the pure motion ``creates'' a universe
with an horizon 
at distance $\Delta x = v \Delta t$ from the observer. As seen from the
rest frame, this part of the physical system does not exist. The ``classical''
entropy of this region is given by the one of its dominant configuration, i.e.
it corresponds to the entropy of a black hole of radius $\Delta x$.} 
\label{v-universe}
\end{figure}

Let us for a moment abstract from the fact that the actual
configuration of the universe implied by \ref{ZPsi} at any time 
describes a curved space. In other words, let's neglect the so called
``cosmological term''. This approximation can make sense at large $N$,
as is the case of the present-day physics. This means at large age of the 
universe~\footnote{To make contact with ordinary physics, consider that, once
expressed in units in which the Planck constant and the speed of light are 1,
the present age of the universe is estimated to be of order $10^{31}$,
and the cosmological constant of order $\Lambda \sim 10^{-61}$.
It is precisely its smallness what historically
allowed to introduce special relativity and Lorentz boosts before addressing 
the problem of the cosmological constant.}.
Let us also assume we can just focus our attention on two observers sitting
on two \underline{inertial} frames, $A$ and $A^{\prime}$, 
moving at relative speed $v$, neglecting everything else.  
For what said above, $v < 1$. 
An experiment is the measurement of some event that, owing to the
fact that happening of something means changing of entropy and therefore
is equivalent to a time progress, 
is perceived as having taken place during
a certain interval of time. Let us
consider an experiment, i.e. the detection of some event, taking place in
the co-moving frame of $A^{\prime}$, as reported by both
the observer at rest in $A$,
and the one at rest in $A^{\prime}$ (from now on we will
indicate with $A$, and $A^{\prime}$, indifferently the frame as well as
the respective observer). Let's assume
we can neglect the space distance separating the two observers, or suppose
there is no distance between them \footnote{In our scenario, 
huge (=cosmic) distances
have effect on the measurement of masses and couplings.}. 
For what we said above, such a detection amounts in observing
the increase of entropy corresponding to the occurring 
of the event, as seen from $A$, 
and from $A^{\prime}$ itself. Since we are talking of the same event, the
\emph{overall} change of entropy will be the same for both 
$A$ and $A^{\prime}$.
One would think there is an ``absolute'' time
interval, related to the evolution of the universe corresponding to the
change of entropy due to the event under consideration. However, 
the story is rather different as soon as we consider \emph{time measurements}
of this event, as reported by the two observers, $A$ and $A^{\prime}$. 
The reason is that the two observers will in general attribute
in a different way what amount of entropy change has to be considered
a change of entropy of the ``internal'' system, and which amount
refers to an ``external'' change. Proper time measurements have to do with
the \emph{internal} change of entropy. 
For instance, consider the entropy of
all the configurations contributing to form, say, a clock. The part of phase
space describing the uniform motion of this clock will not be taken into
account by an observer moving together with the clock, as it will not even
be measurable. This part will however be considered by the other observer.    
Therefore, when reporting measurements of time intervals made by 
two clocks, one co-moving with $A$, and one seen
by $A$ to be at rest in $A^{\prime}$, owing to a different way of
attributing elements within the configurations building up the system,
between ``internal'' and 
``external'', we will have in general two different time
measurements.  
Let us indicate with $\Delta S$ the change of entropy as is observed
by $A$. We can write:
\ba
\Delta S \, (\equiv \Delta S(A) ) 
& = & \Delta S ({\rm internal}\, = \, {\rm at \, rest}) 
\, + \, \Delta S ({\rm external})  \label{DeltaSS} \\
&& \nn \\
& = & \Delta S (A^{\prime}) \, + \, 
\Delta S_{\rm Kinetic}(A) \, ,  
\label{DeltaSSk}
\ea
with the identifications 
$\Delta S ({\rm internal}
\, = \, {\rm at \, rest}) \equiv \Delta S (A^{\prime})$ and
$\Delta S ({\rm external}) \equiv \Delta S_{\rm Kinetic}(A)$.
In section~\ref{eSp} we discussed how the entropy of a three sphere is 
proportional to $N^2 = E^2$. This is therefore also the entropy of 
the average, classical universe, that in the continuum limit, via the 
identification of total energy with time, can be written as: 
\be
S \; \propto \; \left( c {\cal T} \right)^2 \, ,
\label{ScT2}
\ee
where ${\cal T}$ is the age of the universe.
This relation matches with the Hawking's expression of the entropy of a 
black hole of radius $r = c {\cal T}$ \cite{Bardeen:1973gs,Bekenstein:1973ur}.
It is not necessary to write explicitly  the proportionality constant in 
(\ref{ScT2}),
because we are eventually interested only in ratios of entropies.
During the time of an event, $\Delta t$, the age of the universe passes
from ${\cal T}$ to ${\cal T} + \Delta t$, and
the variation of entropy, $\Delta S = S ({\cal T} + \Delta t)-S({\cal T})$,
is:
\be
\Delta S \; \propto \; \left( c \Delta t \right)^2 \, + \, 
c^2 {\cal T}^2 \left( {2 \Delta t \over {\cal T}}  \right) \, . 
\ee 
The first term corresponds to the entropy of a ``small universe'', 
the universe which is ``created'', or ``opens up'' around an observer
during the time of the experiment, and embraces within its horizon the entire 
causal region about the event. The second term is a 
``cosmological'' term, that couples the local physics to the history of the 
universe. The influence of this part of the
universe does not manifest itself through elementary, classical causality
relations within the duration of the event, but indirectly, through a (slow)
time variation of physical parameters such as masses and couplings,
(we refer to \cite{npstrings-2011} for a discussion of the time dependence 
of masses and couplings. See also~\cite{spi}).
In the approximation of our abstraction to the rather ideal case of
two inertial frames, we must neglect this part, concentrating the discussion
to the local physics. In this case, each experiment must be considered
as a ``universe'' in itself.
Let's indicate with $\Delta t$ the time interval as reported by $A$, and 
with $\Delta t^{\prime}$
the time interval reported by $A^{\prime}$.
In units for which $c=1$, and omitting the normalization
constant common to all the expressions like~\ref{ScT2} , 
we can therefore write:
\be
\Delta S(A) \rightarrow \langle \Delta S(A)  \rangle 
\approx (\Delta t)^2 \, ,
\label{dst}
\ee
whereas
\be
\Delta S (A^{\prime}) \rightarrow 
\langle \Delta S (A^{\prime})  \rangle
\approx (\Delta t^{\prime})^2 \, ,
\label{dstp}
\ee
and
\be
\Delta S_{\rm Kinetic} (A) ~ = ~ ( v \, \Delta t )^{2} \, . 
\label{vt2}
\ee
These expressions have the following interpretation. As seen from $A$,
the total increase of entropy corresponds to the black hole-like entropy
of a sphere of radius equivalent to the time duration of the 
experiment. Since $v=c=1$ is the maximal 
``classical'' speed of propagation of information, all the classical 
information about the system is contained within the horizon set by the 
radius $c \Delta t= \Delta t$. However, when $A$ attempts to refer this time 
measurement to what $A^{\prime}$ could observe, it knows that 
$A^{\prime}$ perceives itself at rest, and 
therefore it cannot include in the computation of entropy also the change in 
configuration due to its own motion (here it is essential that we consider 
inertial systems, i.e. constant motions). 
``$A$'' separates therefore its measurement 
into two parts, the ``internal one'', namely the one involving changes that
occur in the configuration as seen at rest by $A^{\prime}$ 
(a typical example is for instance a muon's decay at rest in
$A^{\prime}$), and a part accounting 
for the changes in the configuration due to the very being 
$A^{\prime}$ in motion at speed $v$.
If we subtract the internal changes, namely we think at the system at rest in
$A^{\prime}$ as at a point without meaningful physics 
apart from its motion in space \footnote{No internal physics means
that we also neglect the contribution to the energy, and entropy, due to the
mass.}, the entire information about the change of entropy is contained in 
the ``universe''
given by the sphere enclosing the region of its displacement, 
$v^2 (\Delta t)^2 ~ = ~ \Delta S_{\rm Kinetic} (A)$. In other words,
once subtracted the internal physics, the system behaves, from the point
of view of $A$, as a universe which expands at speed $v$, because the only
thing that happens is the displacement itself, of a point otherwise fixed
in the local universe (see figure~\ref{v-universe}). 
Inserting expressions \ref{dst}--\ref{vt2} in \ref{DeltaSSk} we obtain:
\be
(\Delta t)^2 ~ = ~ { (\Delta t^{\prime})^2 \over 1 - v^2 } \, , 
\ee
that is:
\be
\Delta t ~ = ~ { \Delta t^{\prime} \over \sqrt{1 - v^2}  } \, .
\ee
The time interval as measured by $A$ results to be longer by a factor 
$(\sqrt{1 - v^2})^{-1}$ than as measured by $A^{\prime}$.
In this argument the bound on the speed of information, 
and therefore of light, enters 
when we write the variation of entropy of the ``local
universe'' as $\Delta S = (c \Delta t)^2$. If $c \to \infty$, namely, if within
a finite interval of time an infinitely extended causal region opens up around
the experiment, both $A$ and $A^{\prime}$ turn out to have access to the full
information, and therefore $\Delta t = \Delta t^{\prime}$.
This means that they observe
the same overall variation of entropy.

\subsubsection{the space boost}

In this framework we obtain in quite a natural way the Lorentz 
\underline{time} boost. 
The reason is that, for us, time evolution is directly related to entropy 
change, and we identify configurations (and geometries) through their entropy. 
The space length is somehow a derived quantity, 
and we expect also the space boost to be 
a secondary relation. Indeed, it can be easily derived from the time boost, 
once lengths and their measurements are properly defined. However, these 
quantities are less fundamental, because they are related to the classical 
concept of geometry. 
We could produce here an argument leading to the space boost. However, 
this would basically be a copy of the classical derivation within the 
framework of special relativity. The derivation of the time boost through 
entropy-based arguments opens 
instead new perspectives, allowing to better understand where relativity ends
and quantum physics starts. Or, to better say, it provides us with an 
embedding of this problem into a scenario that contains both these aspects,
relativity and quantization, as 
particular cases, to be dealt with as useful approximations.

\subsection{General time coordinate transformation}
\label{gtrans}

Lorentz boosts are only a particular case of
the general coordinate transformation, obtained
within the context of General Relativity; in that case the measure 
of time lengths 
is given by the time-time component of the metric tensor. In the absence of 
mixing with space boosts, i.e., with a diagonal metric, we have:
\be
(d s)^2 ~ = ~ g_{00} (dt)^2 \, . 
\ee 
As the metric depends on the matter/energy content through the Einstein's 
Equations:
\be
{\cal R}_{\mu \nu} - {1 \over 2} g_{\mu \nu} {\cal R} \, 
= \, 8 \pi G_N T_{\mu \nu} \, ,
\label{Eeq}
\ee
$g_{00}$ can be computed when we know the energy of the system. For instance,
in the case of a particle of mass $m$ moving at constant speed 
$v$ (inertial motion),
the energy, the ``external'' energy, is the kinetic energy 
${1 \over 2} m v^2$, and we recover the
$v^2$-dependence of the Lorentz boost~\footnote{In the determination of the 
geometry, what matters here is not the full force experienced by the particle
but the field  in which the latter moves. The mass $m$ therefore drops out from
the expressions (see for instance~\cite{landau}).}.   

In the simple case of the previous section, we have considered 
the physical system of the wave packet as
decomposed into a part experiencing an ``internal'' physics, and a 
part which corresponds to the point of view
of the center of mass, that is a part 
in which the complex internal physics is dealt with as a point-like 
particle. The Lorentz boost has been derived as the consequence of a 
transformation of entropies. 
Indeed, our coordinate transformation is based on the same physical grounds
as the usual transformation of General Relativity, 
based on a metric derived from the energy tensor. 
Let us consider the transformation
from this point of view:
although imprecise, the approach through the linear approximation
helps to understand where things come from. In the linear
approximation, where one keeps only the first two terms of the expansion of the
square-root $\sqrt{1 - v^2/c^2}$, the Lorentz boost can be obtained from
an effective action in which in the Lagrangian appear the rest and the kinetic
energy. These terms correspond to the two terms on the r.h.s. of 
equation~\ref{DeltaSSk}. 
Entropy has in fact the dimension of an energy multiplied by a time 
\footnote{By definition, $d S = dE / T$, where $T$ is the temperature, and
remember that in the conversion of thermodynamic formulas, 
the temperature is the inverse of time.}. Approximately, we can write:
\be
\Delta S ~ \simeq ~
\Delta E \Delta t \, , 
\label{SEt}
\ee 
where $\Delta E$ is either the kinetic, or the rest energy.
The linear version of the Lorentz boost is obtained by inserting 
in~(\ref{SEt}) the expressions $\Delta E_{rest} = m$ and 
$\Delta E_{kinetic} = {1 \over 2} m v^2$. In this case, the linearization
of entropies lies in the fact that we consider the mass a constant, instead
of being the full energy of the 
``local universe'' contained in a sphere of 
radius $\Delta t$, i.e. the energy (mass) of a black hole of radius 
$\Delta t$: $m = \Delta E = \Delta t / 2$. 
In our theoretical framework, the general expression of the time coordinate
transformation is:
\be
(\Delta t^{\prime})^2 ~ = ~ 
\langle \Delta S^{\prime}(t) \rangle \, - \, 
\langle \Delta S^{\prime}_{external}(t) \rangle \, .
\label{SpSS}
\ee
Here $\Delta S^{\prime}(t)$ is the total variation of entropy of the ``primed''
system as measured in the ``unprimed'' system of coordinates:
$\langle \Delta S^{\prime}(t) \rangle \, = \, (\Delta t)^2$.
We can therefore write expression~\ref{SpSS} as:
\be
(\Delta t^{\prime})^2 ~= ~ \left[ 1 - \mathcal{G}(t) \right] (\Delta t)^2 \, , 
\label{dtdt2}
\ee  
where:
\be
\mathcal{G}(t) ~ \stackrel{\mathrm{def}}{=} ~ 
{\Delta S^{\prime}_{external}(t) \over (\Delta t)^2}   \, . 
\label{Gdef}
\ee
With reference to the ordinary metric tensor $g_{\mu \nu}$, we have:
\be
\mathcal{G}(t) ~ = ~ 1 - g_{00}  \, .
\ee
$\Delta S^{\prime}_{external}(t)$ is the part of change of entropy of 
$A^{\prime}$ referred to by the observer $A$
as something that does not belong to the rest frame of $A^{\prime}$. It can be
the non accelerated motion of $A^{\prime}$, as in the previous example, or more
generally the presence of an external force that produces an acceleration.
Notice that the coordinate transformation \ref{dtdt2} starts with a constant
term,~1: this corresponds to the rest entropy term expressed in
the frame of the observer. For the observer, the new time metric is always
expressed in terms of a deviation from the identity.

By construction, \ref{Gdef} is the ratio between the metric in 
the system which is observed and the metric in the system of the observer.
From such a coordinate transformation we can pass to the metric of
space-time itself, provided we consider the coordinate transformation 
between the metric $g^{\prime}$ of a point in space-time,
and the metric of an observer which lies on 
a flat reference frame, whose metric is expressed in flat coordinates. 
We have then:
\be
1 \, - \, \mathcal{G}(t)  ~ = ~ { g_{00}^{(\prime)} \over g^{(0)}_{00} \, 
= \eta_{00} = 1} \, .
\ee
As soon as this has been clarified, we can drop out the denominator and we 
rename the primed metric as the metric tout court.

\subsection{General Relativity}
\label{genRel}

Once the measurement of lengths
is properly introduced, as derived from a measurement of configurations
along the history of the system, 
it is possible to extend the relations also to the transformation of
space lengths. This gives in general the components of the metric tensor
as functions of entropy and time. In classical terms, whenever this 
reduction is possible, this can be rephrased into a dependence on energy 
(energy density) and time. 
They give therefore a generalized, integrated version
of the Einstein's Equations.
Let's see this for the time component of the metric. We want
to show that the metric $g_{00}$ of the effective space-time corresponds
to the metric of the distribution of energy in the classical space, i.e., 
in the classical limit of effective three-dimensional space as it
arises from \ref{ZPsi}. This will mean
that the geometry of the motion of a particle within this space is the
geometry of the energy distribution. In particular, if the energy is 
distributed according to the geometry of a sphere, so it will be the
geometry of space-time in the sense of General Relativity.
To this regard, we must remember that: 

\noindent
${\rm i})$  All these arguments make only sense in the ``classical
limit'' of our scenario, namely only in an average sense, where 
the universe is dominated by a configuration that can be described in 
classical geometric terms. It is in this limit that the universe
appears as three dimensional. Configurations which
are in general non three-dimensional, non-geometric, possibly tachyonic, and,
in any case, configurations for which General Relativity and Einstein 
Equations don't apply, are covered under the ``un-sharping'' relations of the
Uncertainty Principle.
All of them are collectively treated as ``quantum effects'';

\noindent
${\rm ii})$
In the classical limit, \underline{nothing}
travels at a speed higher than $c$. As during an experiment
no information comes from outside the 
local horizon set by the duration of the experiment itself, to cause
some (classical) effects on it, any consideration
about the entropy of the configuration of the object under consideration
can be made ``local'' (tachyonic effects are taken into account
by quantization). That means, when we consider the motion of an
object along space we can just consider the local entropy, 
which depends on, and is determined by, 
the energy distribution around the object.

Having these considerations in mind,
let us consider the motion of a particle, or, more precisely, a 
non-dispersive wave-packet, in the three-dimensional, classical space. 
Consider to perform a (generally point-wise) coordinate transformation to
a frame in which the metric of the energy distribution external 
to the system intrinsically building the wave packet in itself is flat,
or at least remains constant. As seen from this set of frames, along the motion
there is no change of the (local) entropy around the particle, and the right
hand side of \ref{Gdef} vanishes, implying that also the metric of the motion
itself remains constant (remember that \ref{Gdef} in
this case gives the \emph{ratio} between metrics at different points/times).
This means that the metric of the energy distribution 
and the metric of the motion
are the same, and proves the equivalence of \ref{SpSS} and \ref{Gdef}
with the Einstein's equations \ref{Eeq}.
If on the other hand we keep the frame of the observer fixed,
and we ask ourselves what will be the direction chosen by the particle
in order to decide the steps of its motion, the answer will be:
the particle ``decides'' stepwise to go in the direction that maximizes
the entropy around itself. Let us consider configurations in which 
the only property of particles is their mass
(no other charges), so that entropy is directly related to the ``energy
density'' of the wave packet. In this case,  
between the choice of moving toward 
another particle, or far away, the system will proceed in order to increase
the energy density around the particle. Namely, moving 
the particle toward, rather than away from, the other particle,
in order to include in its horizon also the new system. This is how 
gravitational attraction originates in this theoretical framework.

In order to deal with more complicated cases, such as those in which
particles have properties other than just their mass 
(electro-magnetic/weak/strong charge), we need a more detailed description of 
the phase space. In principle things are the same, but the appropriate
scenario in which all these aspects are taken into account is the one
in which these issues are phrased and addressed
within a context of (quantum) String Theory.
This analysis, first presented in Ref.~\cite{spi}, is discussed in detail 
in~Ref.~\cite{npstrings-2011}.

\subsection{The metric around a black hole}
\label{mbkhl}

Let us consider once more the general expression relating 
the evolution of a system as is seen
by the system itself, indicated with $A^{\prime}$,
and by an external observer, $A$, expressions~\ref{DeltaSS} and \ref{DeltaSSk}.
In the large-scale, \emph{classical} limit, the variations of entropy
$\Delta S(A)$ and $\Delta S (A^{\prime})$ can be written in terms
of time intervals, as in \ref{dst} and \ref{dstp}, in which
$t$ and $t^{\prime}$ are respectively
the time as measured by the observer, and the proper time of the system
$A^{\prime}$.
In this case, as we have seen expression~\ref{DeltaSSk} can be written as
$
(\Delta t^{\prime})^2 =  (\Delta t)^2 \, - \, 
\langle \Delta S^{\prime}_{\rm external}(t) \rangle
$ (see expression~\ref{SpSS}),
and the temporal part of the metric is given by:
\be
g_{00} ~ = ~ 
{\langle \Delta S^{\prime}_{\rm external}(t) \rangle 
\over (\Delta t)^2} \, - \, 1   \, . 
\label{g00}
\ee
As long as we consider systems for which $g_{00}$ is far
from its extremal value, expression \ref{g00} constitutes a good approximation
of the time component of the metric. 
However, a black hole does not fall within the domain of this
approximation.
According to its very (classical) definition, 
the only part we can probe of a black hole is the surface at the horizon.
In the classical limit the metric at this surface vanishes:
$g_{00} \to 0$ (an object falling from outside toward the black hole
appears to take an infinite time in order to reach the surface).
This means,
\be
\langle \Delta S_{\rm external} \rangle
~ \approx ~\propto \, \left( \Delta t  \right)^2 \, . 
\ee 
However, in our set up time is only an average, ``large
scale'' concept, and only in the large scale, classical limit we
can write variations of entropy in terms of progress of a time coordinate
as in \ref{dst} and \ref{dstp}.
The fundamental transformation is the one given in expressions 
\ref{DeltaSS}, \ref{DeltaSSk}, and the term $g_{00}$ has only to be understood
in the sense of:
\be
\Delta S (A^{\prime}) \, \longrightarrow \,  
\langle \Delta S (A^{\prime}) \rangle \, \equiv \, 
\Delta t^{\prime} g_{00} \Delta t^{\prime} \, . 
\label{dsag}
\ee
The apparent vanishing of the metric~\ref{g00}
is due to the fact that we are subtracting contributions from the
first term of the r.h.s. of expression~\ref{DeltaSSk}, namely 
$\Delta S(A^{\prime})$,
and attributing them to the contribution of the
environment, the world external to the system of which we consider the proper 
time, the second term in the r.h.s. of \ref{DeltaSSk}, 
$\Delta S_{\rm external}(A)$. 
Any physical system is given by the superposition of an
infinite number of configurations, of which only the most entropic ones
(those with the highest weight in the phase space) build up
the classical physics, while the more remote ones contribute to 
what we globally call ``quantum effects''. Therefore,
taking out classical terms from
the first term, $\Delta S (A^{\prime})$, the ``proper frame'' term,
means transforming the system the more and more into a ``quantum
system''. In particular, this means that the mean value of whatever observable
of the system will receive the more and more contribution 
by less localized, more exotic, configurations, 
thereby showing an increasing quantum uncertainty.
In particular, the system moves toward configurations 
for which $\Delta x \rightarrow \gg 1 / \Delta p$.
Indeed, one never reaches the condition of vanishing of \ref{dsag}, because,
well before this limit is attained, also the notion itself of space, and time,
and three dimensions, localized object, geometry, etc..., are lost. The
most remote configurations in general do not
describe a universe in a three-dimensional space, and the ``energy'' 
distributions are not even interpretable in terms of ordinary observables.
At the limit in which we reach the surface of the horizon, the black hole will
therefore look like a completely delocalized object~\cite{blackholes-g}.

\subsection{Natural or real numbers?}
\label{NorR}

The approach we are proposing, and the fact that from the collection
of binary codes we arrive to the structures of our
physical world, implies
a question about what is after all the world we experience.
We are used to order our observations according to phenomena that
take place in what we call space-time. An experiment, or, better,
an observation (through an experiment), any perception in itself,
basically consists in realizing that
something has changed: our ``eyes'' have been affected by something, that
we call ``light'', that has changed their configuration (molecular, atomic 
configuration). This light may carry information about changes in our
environment, that we refer either to gravitational phenomena, or 
to electromagnetic ones, and so on...
In order to explain them we introduce energies, momenta, ``forces'', i.e.
interactions, and therefore we speak in terms of masses, couplings etc...
However, all in all, what all these concepts refer to is a change in
the ``geometry'' of our environment, a change that ``propagates'' to us,
and eventually results in a change in our brain, the ``observer''.
But what is after
all geometry, other than a way of saying that, by moving along
a path in space, we will encounter or not some modifications? 
Assigning a ``geometry''
is a way of parametrizing modifications.
Is it possible then to invert the logical ordering from reality to
its description? Namely, can we argue that
what we interpret as energy,
or geometry, is simply a code of information \footnote{See for instance 
the ``it from bit'' of J. A. Wheeler, and the work of C. F. Weizs\"{a}cker.}. 
Something happens, i.e. time passes, when some code changes. 
Viewed in this way, it is not a matter of mapping
physical degrees of freedom into a language of abstract codes, but 
the other way around, namely: perhaps
the deepest reality \emph{is} ``information'', that we
arrange in terms of geometries, energies, particles, fields, and
interactions.  
When we ``see'' the universe,
we \emph{interpret} the codes in terms of maps, 
from a space of ``energies'' to a target space, that take the ``shape''
of what we observe as the physical reality. From this
point of view, information is not just something that transmits knowledge
about what exists, but it is itself the essence of what exists, and the 
rationale of the universe is precisely that it ultimately
is the whole of rationale. The quantum (in the sense of indeterministic)
nature of the universe is then the consequence of being any
observable not just a code but a collection, a superposition, of codes.

Reducing everything to a collection of binary codes means reducing
everything to a discrete description in terms of natural numbers,
i.e. to saying that the whole of rationale is numerable.
One may wonder whether natural numbers are enough to encode \emph{all} 
the information of the universe.
At first sight, one would say that real numbers say ``more'', allow to express
more information. Moreover, they appear to be ``real'' in the true sense
of something existing in nature. For instance, one can think to draw with
the pencil a circle and a diameter. Then, one has \emph{physically} realized
two lines of lengths that don't stay in a ratio expressible as a rational 
number. However here the point is: what is really about the microscopical 
nature of these two drawings? At the microscopical level, at the
scale of the Planck length, the notion of space itself is so fuzzy 
to be practically lost. In our scenario, 
an analysis of the superposition
of configurations tells us that, before reaching this scale, remote
configurations, whose contribution is usually collected 
under the Heisenberg's Uncertainty, count more and more. In other words,
the world is no more classical but deeply quantum mechanical, to the point
that the uncertainty in the length of the two lines doesn't allow us
to know whether their ratio is a real or a rational number. In this sense,
this analysis provides further support to an old idea which goes back to 
Konrad Zuse, that all the information of the universe is expressible
through natural numbers, and, as a consequence, the discrete description
of the universe, and in particular of space-time, is not just
an approximation, but indeed the most
fundamental one can think about. After all,
real numbers are introduced in mathematics through definitions
and procedures, whose informational content
can be ``written'' as a text with a computer program. This means that, 
\emph{as a matter of pure information content}, 
real numbers can be introduced via
natural numbers.

\newpage

\providecommand{\href}[2]{#2}\begingroup\raggedright\endgroup


\begin{thebibliography}{10}

\bibitem{assiom-2011}
A.~Gregori, \emph{{Combinatorics, observables, and String Theory},} arXiv
  e-prints (2011) \href{http://www.arXiv.org/abs/arXiv:1103.4000}{{\tt
  arXiv:1103.4000}}.

\bibitem{npstrings-2011}
A.~Gregori, \emph{{Combinatorics, observables, and String Theory, part II},}
  arXiv e-prints (2011) \href{http://www.arXiv.org/abs/arXiv:1103.3998}{{\tt
  arXiv:1103.3998}}.

\bibitem{spc-gregori}
A.~{Gregori}, \emph{{On the Critical Temperatures of Superconductors: a Quantum
  Gravity Approach},} arXiv e-prints (July, 2010)
  \href{http://www.arXiv.org/abs/arXiv:1007.3731}{{\tt arXiv:1007.3731}}.

\bibitem{blackholes-g}
A.~{Gregori}, \emph{{Is the Universe the only existing Black Hole?},} ArXiv
  e-prints (June, 2010) \href{http://www.arXiv.org/abs/1006.5826}{{\tt
  1006.5826}}.

\bibitem{paleo}
A.~Gregori, \emph{A note on the phases of natural evolution,}
\href{http://www.arXiv.org/abs/arXiv:0712.0074 [physics.gen-ph]}{{\tt
  arXiv:0712.0074 [physics.gen-ph]}}.

\bibitem{Rabinowitz:2001ag}
M.~Rabinowitz, \emph{n-dimensional gravity: Little black holes, dark matter,
  and ball lightning,} Int. J. Theor. Phys. {\bf 40} (2001) 875--901,
\href{http://www.arXiv.org/abs/astro-ph/0104026}{{\tt astro-ph/0104026}}.

\bibitem{Bardeen:1973gs}
J.~M. Bardeen, B.~Carter, and S.~W. Hawking, \emph{The Four laws of black hole
  mechanics,} Commun. Math. Phys. {\bf 31} (1973)
161--170.

\bibitem{Bekenstein:1973ur}
J.~D. Bekenstein, \emph{Black holes and entropy,} Phys. Rev. {\bf D7} (1973)
2333--2346.

\bibitem{spi}
A.~Gregori, \emph{An entropy-weighted sum over non-perturbative vacua,}
\href{http://www.arXiv.org/abs/arXiv:0705.1130 [hep-th]}{{\tt arXiv:0705.1130
  [hep-th]}}.

\bibitem{landau}
L.~D. Landau and E.~M. Lifshitz, {\em The Classical Theory of Fields}.

\end{thebibliography}
\end{document}